\documentclass{article}

\usepackage{PRIMEarxiv}

\usepackage{amsmath}
\usepackage{silence}
\usepackage{graphicx}
\usepackage{algorithm}
\usepackage{algpseudocode}
\usepackage{multirow}
\usepackage{url}
\usepackage{siunitx} % for better alignment of numbers

\usepackage{amssymb}

\usepackage{subcaption}
\usepackage[usenames,dvipsnames]{color}
\graphicspath{ {./images/} }
\usepackage[inkscapeformat=png]{svg}
\usepackage{booktabs}

\usepackage{pdflscape} % For landscape pages
\usepackage{longtable} % For tables that span multiple pages
\usepackage{rotating}
\usepackage{tabulary}
\usepackage{adjustbox}

\usepackage{svg}
\usepackage{hyperref}
\title{DQN-Scheduler: A Multi-Objective Optimization Framework for Scheduling Microservices in Cloud Computing
}

\author{
  Abdullah Alelyani \\
  The University of Western Australia \\
  \texttt{abdullah.alelyani@research.uwa.edu.au} \\
   \And
  Amitava Data \\
  The University of Western Australia \\
  \texttt{amitava.datta @uwa.edu.au} \\
  \AND
  Ghulam  Mubasher \\
  The University of Western Australia \\
 \texttt{ghulam.hassan@uwa.edu.au} \\
}

\begin{document}
\maketitle

\begin{abstract}
Cloud computing has emerged as an information technology solution, providing software and infrastructure solutions for companies and individuals. The pay-as-you-go approach has increased demands for the cloud. The massive range of resources, the variety of services, and flexible pricing grab attention. In addition, microservices have emerged as a new way of building software, with applications developed as loosely dependent tasks. Additionally, container technology has boosted the popularity of microservices by offering a platform for this type of architecture. Containers and microservices improve the flexibility and scalability of cloud applications. There are two primary types of microservices: batch and online services, with the majority of applications falling into the online service category. Scheduling microservices is challenging because it requires careful management of resource utilization, load balancing, network latency, reliability, and availability. In this study, we introduce the DQN-Scheduler, a novel reinforcement learning-based agent designed to optimize microservice scheduling in cloud environments. Our approach aims to optimize multiple scheduling objectives simultaneously, such as resource utilization, load balancing, latency, reliability, and availability. To our knowledge, this is the first framework to address all these objectives simultaneously. The DQN-Scheduler was tested against benchmark algorithms in the field. The experimental results demonstrate that the DQN-Scheduler outperforms benchmark algorithms.

\end{abstract}

% }
\maketitle
% \IEEEdisplaynontitleabstractindextext

% \IEEEpeerreviewmaketitle

% \IEEEraisesectionheading{\section{Introduction}\label{sec:introduction}}
\section{Introduction}\label{sec:introduction}
% \IEEEPARstart{C}{loud} 

Cloud computing benefits from the microservice architecture in developing applications. Several reasons for the cloud recently transitioning from developing applications as monolithic architecture to microservices include simpler development and efficient deployment\cite{8362750}. Since each microservice is responsible for performing one task, microservices allow for more accurate and easier development of applications. Loosely coupled microservices shift the requirement of using a single programming language for the entire application to use different languages for different microservices. Two of the most important advantages of microservices are their compatibility with containers for cloud applications\cite{8362750} and their ability to be distributed across cloud physical machines (PMs). Microservices utilize remote procedure calls (RPC) or a RESTful API \cite{zhang2023high} to communicate between them.

Communication between microservices can affect the quality of service (QoS) of the cloud platform by decreasing throughput and increasing latency\cite{8362750}\cite{konjaang2021multi}.In addition, balancing resource utilization contributes to improving energy efficiency. With microservices, balancing the load between resources becomes challenging, as some microservices are either CPU-intensive or memory-intensive. For example, if there are $n$ memory-intensive microservices distributed across $m$ PMs, the memory of the PMs would be heavily used, while the CPUs would remain underutilized, leading to CPU wastage.

Scheduling microservices across PMs while maintaining multiple objectives is challenging. For example, minimizing energy consumption while maximizing resource utilization can lead to SLA violations if the CPU of PMs is heavily utilized. Trade-offs between conflicting objectives are a common strategy in the scheduling process. However, these trade-offs do not guarantee optimal scheduling, especially in dynamic environments like cloud computing.

%Additionally, the high volume of requests for accessing services on the cloud raises the need to scale microservices applications. However, scaling microservices can lead to an imbalance in resource utilization. In addition, it leads to the waste of cloud resources. Scaling microservices applications increases the availability of applications, but it requires a strategy that helps improve resource efficiency and reliability. For example, assume that an application has $m$ microservices, with $m_1$ being the entry service that handles all client requests and $m_n$ being the microservice responsible for handling logout operations. It is wise to scale $m_1$ since it manages a large number of requests. On the other hand, $m_n$ might not require as much scaling as $m_1$. Scaling without considering the load on the cloud can lead to resource waste. Moreover, scaling microservices within the same PM increases the risk of single points of failure, causing the application to become unavailable.}

Resource utilization, throughput, availability, reliability, and load balancing are among the main objectives for enhancing cloud performance. However, these objectives pose challenges in developing cloud scheduling strategies, as they often conflict. Enhancing one objective may degrade others. This motivates us to propose a framework to find the optimal solution for enhancing resource utilization, balancing the load between resources within each PM and among PMs, reducing latency, and enhancing the availability and reliability of cloud computing applications. Our main contributions are as follows:
\begin{enumerate}
 \item \textbf{Optimizing Microservices Scheduling Framework:} We use a double-deep Q-Network agent model. The agent is trained to provide scheduling decisions that consider the conflicting objectives mentioned above.
 \item \textbf{Developing a Reward Model:} This model utilizes the Pareto Front approach to determine the agent's reward during training based on the optimal solution for conflicting objectives. It aims to guide the agent's behaviour to improve scheduling decisions.
 \item \textbf{Developing an Action Selection Model:} This model selects the best action provided by the agent based on real-time environmental conditions. Given the stochastic nature of the cloud environment, this selection approach employs a Bayesian inference model.
\item \textbf{Developing Load Balancing and QoS Models:} We designed models to enhance the scheduling strategy by monitoring resource utilization, load balancing, latency,  reliability and availability of microservices.
 \item \textbf{Conducting Extensive Experiments:} We conducted experiments that assessed the efficiency of our proposed framework, including comparisons with benchmark algorithms such as the Genetic Algorithm (GA), Particle Swarm Optimization (PSO), and Best-Fit Algorithm (BFA).
\end{enumerate}

The rest of our research is organized as follows: Section 2 reviews related work; Section 3 describes the proposed system models; Section 4 describes the architecture of DQN-Scheduler; Section 5 discusses the cloud environment scenario and workload; Section 6 introduces the benchmark algorithms used for comparison with our proposed framework; Section 7 details the experiment setup and results; and Section 8 provides the conclusion.

\section{Current Studies}

In this section, we discuss current studies that proposed to address scheduling with multi-objectives in detail. Studies included in this part of our research must meet the following criteria: 
\begin{enumerate}
   \item Scheduling microservices to cloud considering multi-objectives that are conflicted.
    \item Considering real-time scheduling scenarios.
    \item Considering the dynamic environment of cloud computing.
    \item Examining the proposed approach using real-world datasets or implementing the approach on an actual cloud computing platform. 
    \item Considering the significant volume of tasks arriving on the cloud platform during the experiment.
\end{enumerate}

We decided to limit the included studies to those meeting the criteria above to ensure the studies best mimic the reality of cloud computing platforms. Furthermore, we aim to narrow our search to strategies that utilize reinforcement learning models, like deep Q-learning.

A multi-objective approach that aims to minimize the communication overhead between microservices and balance the load (RSDQL) is proposed in \cite{lv2022microservice}. RSDQL utilized a deep Q-learning algorithm to enhance the development strategy and balance the load among resources. The approach is modelled as a multi-objective problem. It primarily considers the communication overhead. Additionally, the model aims to avoid degrading the performance of scheduled task execution by considering the capacity of nodes. An undirected and weighted graph is used to demonstrate the communication overhead between microservices. Additionally, the proposed model uses the graph to minimize the interaction overhead between microservices and to reduce the variance of utilizing resources to balance the load. Furthermore, an elastic scaling approach was introduced to increase the number of deployed instances hosting new microservices and manage the dynamic resource utilization of applications. Elastic scaling also aims to manage QoS by utilizing real-time resource monitoring utilities, and a resource utilization threshold. The experimental results demonstrate that RSDQL was able to reduce the average response time, balance loads, and improve scalability. However, energy consumption and the limitations of edge resources were not taken into consideration.

An online microservices orchestration for managing Internet of Things (IoT) application allocation is proposed in\cite{9723469}. The infrastructure of the orchestration is based on a policy-based DRL algorithm. The study aims to address application allocation with two objectives: minimizing the long-term energy consumption and the end-to-end execution time of the deployed application. The allocation problem was considered a multi-objective Markov decision process (MOMDP). As a result, a model named MOTION was proposed as a multi-objective deep reinforcement learning model. MOTION aims to optimize the dynamic orchestration of allocating IoT microservices in cloud environments. The proposed model was tested against the proposed baseline algorithms, showing the efficiency of the proposed model in reducing long-term energy consumption and execution time for applications. Memory and I/O allocation are not considered in this approach. The real-world scenario of microservices communication is not considered as well.

A framework employing deep reinforcement learning (DRL) to limit incoming requests to containers in Alibaba's data centre (Noah) is proposed in \cite{10098822}. The framework is an adaptive limiter for the large-scale microservice platform running in containers. Noah comprises three main components: a limiter, a monitoring component, and a DRL-based decision-maker. It is a closed-loop system where requests first arrive at the limiter. Then, based on a threshold, requests are either forwarded to the container or rejected. The monitoring components retrieve the statuses of containers and provide feedback to the DRL-based decision-maker. The DRL model is responsible for setting up the threshold for the limiter in real-time. Four goals were introduced for the framework: 1) the limiter should automatically adapt without domain knowledge; 2) the framework must be available all the time; 3) the framework should react in time based on the load situation; and 4) the framework should minimize the number of rejected requests. The DRL is also trained with synthetic data representing extreme situations to ensure the model can handle unexpected requests. Noah shows an efficient adjustment to the threshold in real-time based on the load of the containers. Noah accounts for a limited number of scenarios. However, there is no evidence of Noah's ability to achieve considerable performance in the dynamic environment of cloud computing.

A framework for addressing the issues of multi-dependencies between microservices in edge computing (GRLD) is proposed in \cite{10162207}. Microservice architecture (MSA) involves multi-dependencies that represent a communication sequence between microservices, called a call graph. The proposed framework utilizes a graph convolutional network (GCN) to extract features from the dependencies between the microservices. GRLD then passes this feature vector to a decision-making process. The decision-making process employs a DRL model that deploys microservices to the edges. The main goal of GRLD is to incorporate multiple pieces of information (e.g., multi-call-graph scenarios, resource utilization, edge node situations) and allow the DRL agent to make decisions about the deployment process. In addition, the agent tries to meet QoS parameters. The experimental results demonstrate that GRLD exhibits minimal deployment overhead costs compared to baseline algorithms. The results demonstrate the efficiency of maintaining QoS parameters. However, the model was not generalized to adapt to real-world scenarios.

An artificial intelligence algorithm utilizing deep Q-learning (DQTS) is proposed in \cite{tong2020scheduling}. It aims to schedule tasks, considering microservice dependencies. DQTS considers microservice dependencies as directed acyclic graph (DAG). The proposed algorithm was developed to minimize makespan and load balancing. The scheduler consists of three components: 1) the resource dynamic allocator, which adjusts and refreshes the available resources; 2) the task allocation controller, which allocates the different types of tasks; and 3) the deep Q-learning algorithm. The deep Q-learning combines a deep neural network and a Q-learning algorithm to provide the scheduling decisions. The experimental results show that DQTS improves cloud performance. Additionally, DQTS achieves the minimum makespan and the best load balance. However, DQTS was evaluated against a scientific workflow benchmark, which does not include real-world scenarios and does not account for crucial objectives like energy efficiency.

A cloud scheduling framework leveraging queuing theory and reinforcement learning to optimize energy consumption and resource utilization (QEEC) is proposed in \cite{ding2020q}. The QEEC consists of two phases: a centralized task dispatcher and a scheduler. The dispatcher monitors the user requests by utilizing a global request queue on the cloud side. The queue serves as a buffer for incoming requests. The scheduler stores incoming requests in local request queues on each node. It then reorders requests based on QoS and distributes them among nodes to minimize overall CPU utilization and SLA violations. The Q-learning-based scheduler treats each node separately during the scheduling process. QEEC minimizes task execution time by considering task lifetime. Extensive experiments utilizing the M/M/S queuing system were conducted to evaluate QEEC. The experimental results show that QEEC efficiently reduces energy consumption and minimizes average response time. However, the framework considered a limited number of nodes that do not reflect the cloud environment's reality.

An algorithm for enhancing the makespan in cloud computing utilizing a Q-learning algorithm (QL-HEFT) is proposed in \cite{tong2020ql}. QL-HEFT aims to enhance the overall performance of scheduling dependent tasks into cloud resources. The algorithm consists of a data centre broker and a task scheduler. The broker distributes the tasks to the cloud resources, considering QoS parameters. QL-HEFT presents dependent tasks as a DAG graph. Additionally, the HEFT algorithm is employed in this research to reduce the makespan. Two phases that construct the HEFT algorithm include task prioritizing and processor selection phases. In the task prioritizing phase, tasks are ranked based on their rank value. In the second phase, tasks are scheduled into the processor, considering the earliest completion time strategy. Furthermore, a Q-learning algorithm is employed to make decisions during scheduling tasks. The main goal is to reduce the makespan and response time. An experiment was conducted to evaluate QL-HEFT and compare its performance with benchmark algorithms such as HEFT\_D, HEFT\_U and CPOP. The experimental results demonstrate that QL-HEFT outperforms the algorithms selected in the performance comparison. However,  QL-HEFT has limitations in solving large-scale tasks because the Q-table becomes significantly large and expensive to update.

A method proposed to address two objectives: makespan and resource utilization (BCRN) is proposed in\cite{asghari2024bi}. The authors claimed that employing standard learning methods to solve scheduling problems requires numerous transitions to converge to the optimal solution. Therefore, BCRN aims to tackle these issues utilizing a Pareto-based algorithm and a learning-based population. In addition, BCRN aims to address the issues of scheduling dependent tasks. BCRN comprises two stages: the learning and the selection stages. In the learning stage, two Q-learning agents are trained to reduce makespan and enhance resource utilization. In the selection stage, NSGA-3 is utilized to identify the best solutions that meet the research objectives. The experimental results demonstrate that BCRN effectively reduces makespan and enhances resource utilization of the cloud. However, BCRN was evaluated using a scientific workflow benchmark, which does not include real-world scenarios. In addition, more agents are needed to enhance the learning process.

An approach utilizing a deep Q-learning model with multiple DVFS (dynamic voltage and frequency scaling) algorithms (DDQ-EES) is proposed in \cite{zhang2018double}. DDQ-EES aims to address energy consumption in data centres and selects the best DVFS algorithm to execute tasks, considering their deadlines. It employs a double-deep Q-learning model consisting of two deep Q-network models. The first model computes Q-values for each DVFS, while the second model tunes the parameters of the generated Q-network. DDQ-EES utilizes a double-deep Q-learning model to maintain model stability. The model adopts the rectified linear units (ReLU) function instead of the sigmoid function to prevent gradients from vanishing. The experimental results demonstrate that DDQ-EES reduces average energy consumption. Furthermore, the results indicate that QQL-EES achieves higher training efficiency. However, DDQ-EES did not employ explore-exploit strategies during the training stage. Additionally, it was not considered for real-world scenarios.

An online resource scheduling framework that utilizes a deep Q-learning algorithm is proposed in \cite{peng2020multi}. The framework aims to address the challenge of decreasing energy consumption and maintaining the QoS. The framework aims to solve the conflict objectives using the Q-learning algorithm. The framework is composed of three layers, including the workload layer, scheduling control layer, and data centre layer. In the workload layer, the users' requests are classified. Then the framework assigns the tasks to a waiting list to be executed, considering the dependencies between tasks. In the scheduling control layer, the tasks are scheduled into the cloud resources. This layer consists of a task monitor, resource monitor, scheduling policy model, energy consumption model, SLA, and scheduling process. The scheduling policy model is a DQN agent that provides optimal scheduling decisions. The agent consists of two deep neural networks that learn simultaneously to enhance the learning process. The experimental results demonstrate that the framework provides trade-off between energy consumption and task makespan by adjusting the weight of rewards. However, the framework is limited when dealing with large-scale tasks. It would also be costly and complex when handling constraints between various resources in the cloud system.

An approach that aims to minimize execution time and energy consumption (E-AEO-AOA) is proposed in \cite{yeganeh2023novel}. The approach  utilizes a combination of optimization algorithms, including artificial ecosystem-based optimization (AEO) and arithmetic optimization algorithm (AOA). The E-AEO-AOA aims to address the issue of offloading tasks from mobile devices to fog or the cloud, considering task execution time and energy consumption. The E-AEO-AOA queues tasks that may be offloaded. Then E-AEO-AOA decides which tasks will be offloaded. AEO and AOA search for the optimum solution for offloading tasks. Additionally, AEO and AOA are discretized utilizing the round function. A Q-learning model is utilized to hybridize the optimization algorithms (AEO and AOA). The experimental results demonstrate that E-AEO-AOA outperforms the competitor algorithms in most cases. However, the approach demonstrates significant complexity and requires more memory and time.

A weighted double deep Q-learning model based on a reinforcement learning algorithm (WDDQN-RL) is proposed in \cite{li2022weighted}. The WDDQN-RL aims to minimize makespan and costs in cloud platforms. Two levels of scheduling strategy are introduced in the study. Each level of the strategy consists of an agent. The agent at the first level enhances the task scheduling order, considering task dependencies. The agent at the second level optimizes resource allocation. Furthermore, time and cost are optimized by the first and second levels of scheduling, respectively. A dynamic sensing mechanism (DSM) adjusts the approach's attention to one of the optimization objectives. The experimental results demonstrate the superiority of WDDQN-RL in various aspects, including solution quality, diversity, and running time. However, WDDQN-RL does not account for real-world scenarios and was not evaluated against different resource constraints.

 Current studies neglect the stochastic nature and real-time changes in the cloud environment. They also overlook the fluctuating resource utilization during scheduling, a key characteristic of the cloud. Additionally, the geographical distribution of data centers (zones) is ignored, resulting in overlooked latency of communications between application tasks. Considering these zones during scheduling is crucial for improving reliability and availability and for simulating real-world scenarios. Moreover, current studies do not address multiple conflicting objectives, ignoring the complex nature of the cloud environment. Table \ref{tab:RL_Comparison} summarizes the key features of current studies, goals, multiple conflicting objectives, methods, experimental results, and limitations.
\begin{table*}
    \centering
    \resizebox{\textwidth}{!}{%
     
     \footnotesize
    \begin{tabular}{|p{2.3cm}|p{2.3cm}|p{.5cm}|p{3cm}|p{4.3cm}|p{4.3cm}|p{4.3cm}}
        \hline
        \textbf{Approach and Key Features} & \textbf{Goals} &  \textbf{MCO}&\textbf{Methodologies} & \textbf{Experimental Results}&\textbf{Limitations} \\ \hline
        RSDQL\cite{lv2022microservice} Utilizes deep Q-learning-based multi-objective approach & Minimize communication overhead and balance load among microservices &  X & RSDQL agents with deep Q-learning algorithm, undirected and weighted graph to represent microservice communication & Reduced average response time, balanced loads, improved scalability & Did not accounting for energy consumption and resource limitations in edge computing.\\ \hline

        \cite{9723469} is a policy DRL-based orchestration & Minimize energy consumption and end-to-end execution time for IoT applications & X &Multi-objective deep reinforcement learning, multi-objective Markov decision process & Reduced long-term energy consumption, improved end-to-end execution time & Did not consider memory and I/O allocation for microservices, nor does it account for real-world scenarios in terms of the interconnections between microservices.\\ \hline

        Noah \cite{10098822}  proposes a deep reinforcement learning-based adaptive limiter for large-scale microservices & Limits incoming requests to containers in data centers & X & Limiter, monitoring component, and DRL-based decision-maker & Minimized number of rejected requests, automatic adaptation without domain knowledge  &It adapts to a specific scenario and does not consider the dynamic environment.  \\ \hline
        
        GRLD\cite{10162207} is a Graph Convolutional Network (GCN)-based approach & Address multi-dependencies between microservices in edge computing to maintain QoS &  X &GCN to extract features from call-graph, DRL model for decision-making & Minimal deployment overhead cost, efficient QoS maintenance  & The model was not generalized to adapt to real-world scenarios.\\ \hline
        
        DQTS\cite{tong2020scheduling}  utilizes a deep Q-learning-based scheduling for cloud resources & Minimizes makespan and maintains load balancing &  X &DAG-based task dependencies, deep Q-learning with deep neural network &  Outperformed other algorithms in terms of makespan and load balancing & Tested against a scientific workflow benchmark, which does not include real-world scenarios and does not account for crucial objectives like energy efficiency  \\ \hline
        
        QEEC\cite{ding2020q}  proposes Q-learning-based cloud scheduling framework & Optimize energy consumption and resource utilization &  X &Queuing theory with Q-learning-based scheduler, centralized task dispatcher, scheduler on each node & Reduced energy consumption and minimized average response time  & Did not evaluated agonist large number of nodes that reflecting the reality of the cloud environment.  \\ \hline

        QL-HEFT\cite{tong2020ql}  utilizes Q-learning-based algorithm for cloud computing scheduling & Reduce makespan and optimize overall scheduling performance &  X &Q-learning with HEFT algorithm, datacenter broker, task scheduler &  Outperformed other benchmark algorithms in performance and makespan reduction  & Not adaptable to large-scale tasks and would be time-consuming for updating\\ \hline

        BCRN\cite{asghari2024bi} is a multi-agent bi-objective scheduling using Q-learning and NSGA-3 & Address makespan and resource utilization in cloud environments &  X &Pareto-based algorithm, two Q-learning agents, NSGA-3 for solution selection & Effective makespan reduction and enhanced resource utilization  &Tested against a scientific workflow benchmarks, which does not include real-world scenarios,  more agents are needed to enhance the learning process \\ \hline
        
        DDQ-EES\cite{zhang2018double} is a Deep Q-learning-based approach for energy efficiency in data centers & Reduce energy consumption using multiple DVFS algorithms &  X &Double-deep Q-learning model, rectified linear units (ReLU) for stability & Reduced energy consumption and improved training efficiency & Did not employ explore-exploit strategies during the training stage. Additionally, it was not considered for real-world scenarios\\ \hline
        
        DQN Framework \cite{peng2020multi} is a Deep Q-learning-based cloud scheduling framework & Optimize energy consumption and task makespan &  X &Multi-layered architecture, DQN agent for optimal scheduling decisions & Trade-off between energy consumption and makespan based on reward weighting  & It is limited when dealing with large-scale tasks, and it is costly and complex when handling constraints between various resources in the cloud system \\ \hline

        E-AEO-AOA\cite{yeganeh2023novel}  is a combination optimization algorithm for cloud resource management & Minimize execution time and energy consumption &  X &Artificial Ecosystem-based Optimization (AEO) and Arithmetic Optimization Algorithm (AOA), Q-learning for hybridization & Outperformed competitor algorithms in most cases  & It is complex and expensive due to the need for more time and memory. \\ \hline
        
        WDDQN-RL\cite{li2022weighted}  is a reinforcement learning-based cloud platform optimization & Minimize makespan and costs & X & Weighted double deep Q-learning, Dynamic Sensing Mechanism (DSM) for attention adjustment & Superior solution quality, diversity, and running time  & Did not account for real-world scenarios and was not evaluated against different resource constraints\\ \hline
    \end{tabular}}
    \caption{Comparison of Related Work Approaches for Scheduling, Resource Management, and Load Balancing in Cloud Environments. Where MCO stands for multiple conflicting objectives.}
    \label{tab:RL_Comparison}
\end{table*}

\section{System Model}\label{system_model}

In this section, the system models are introduced. They encompass load balancing, quality of service (QoS), latency, system architecture, and multi-objective models. We aim to provide a deep understanding of scheduling microservices processes and the issues of scheduling microservices to the cloud. additionally, we will illustrate the methodology employed and discuss the metrics utilized to measure the performance of our framework in this section.

\subsection{Cloud Platform Model}

Cloud platforms are distributed worldwide. With the Internet of Things (IoT) coming into the picture, having data zones close to clients becomes ideal. A data center is denoted as $D$. $D$ consists of a set of data zones, represented as $D$=$\{z_1, z_2...z_z\}$. These zones are interconnected with a high-speed network to mitigate latency. The zones comprise physical machines (PMs) with heterogeneous resources such as CPUs, memory capacities, and network bandwidth in clusters. Additionally, each cluster consists of a limited number of resources. Clusters are managed by cluster orchestration such as Kubernetes and Docker daemon. In our work, we focus on Kubernetes technology for cluster orchestration.

Cluster architecture managed by Kubernetes comprises three main components: node, pod, and container. A node can be a virtual machine (VM) with fixed resources (e.g., CPU, memory, bandwidth) or a physical machine like a bare metal machine. The node provides an isolated environment for running applications. A cluster consists of several nodes denoted as $C$=$\{n_1,n_2....n_c\}$. A set of pods, denoted as $P={p_1, p_2, ... p_p}$, is allocated to a node, where $P \in C$. A pod is the smallest unit of the orchestration architecture, serving as a computational unit. Each pod consists of a group of containers, denoted as $R \in P$, and $R$=$\{r_1,r2.....r_r\}$ where containers handle the tasks of running applications.

\subsection{Microservies Model}

Two architectures shape the structure of applications: monolithic and microservices. The monolithic architecture builds the application as a single unit. However, microservices promote the concept of loosely coupled tasks that construct the application. With the rise of IoT applications, microservices emerge as the preferred choice for developing the applications. Furthermore, edge applications demand high computational resources utilizing microservices architecture. Edge tasks (only CPU-intensive) can be offloaded to cloud resources, while the remaining application tasks can be processed on the edge device. In such scenarios, microservices offer advantages by reducing energy consumption, improving communication during offloading, and becoming cost-efficient.

Microservices are hosted by $n$ containers and distributed across data centers. Therefore, communication between dependent tasks is crucial. Loosely coupled communication tasks are presented as a graph where each edge represents a microservice and vertices represent communication between microservices, denoted as $G=\{E,V\}$. Microservices provide a sophisticated platform to deploy applications, such as IoT applications, with a high level of reliability, scalability, and availability. Quality of service (QoS) parameters such as service level agreement (SLA), and replica of the servers are mostly specified by clients during the submission of the requests to deploy the application. We consider the number of replicas of each microservice to be specified during scheduling. Thus, each microservice $M$ can be composed of several replicas, denoted as $M_{replica}^m=E_p$ where $E_p$ is a constant representing the number of replicas.

\subsection{Multi-objective Model}
This subsection provides a brief overview of multi-objective model management. We use Pareto front optimization, which is detailed as follows: 

Multi-objective optimization is a key focus of our research to enhance the reliability of cloud computing systems.  Effective resource utilization management is essential to minimize resource waste and optimize system performance. Network traffic costs also impact the performance of the cloud. Hence, our research targets reducing network traffic by minimizing communication between dependent microservices across different geographical zones. Achieving high reliability requires a sophisticated scheduling approach to manage service availability and mitigate the risk of a single point of failure. A single point of failure risk arises when all service replicas from one application are housed in a single PM. Such circumstances produce a high probability of being unavailable. Additionally, the intensive node resource utilization directly impacts reliability.

However, these objectives are conflicted, making it challenging to improve them simultaneously. To address this challenge, we employ Pareto Front optimization in our research. Pareto Front optimization balances multiple conflicting objectives.

Multi-objective optimization problems (MOPs) involve optimizing two or more conflicting objectives simultaneously. The optimal solution to such problems often involves a trade-off between the objectives \cite{9817389}. However, we aim to find the optimal point of the trade-off between these objectives. For example, we aim to reduce network traffic by consolidating microservices geographically. However, this consolidation may affect system load balancing and reliability.

To address these challenges and enhance microservice scheduling performance, we propose a multi-objective optimization approach with $O_d$ objectives utilizing Pareto front (PF) optimization. PF enables us to explore the solution space and find optimal solutions for all objectives. In our case, the solution space is defined by $O_d-1$\cite{9817389}. The MOPs can be formulated as\cite{li2021estimation}:

\begin{equation}\label{EQ:9}
\begin{split}  
\text{min \ }{\mathbf{x}} \quad f(\mathbf{x}) = \left[ f_1(\mathbf{x}), f_2(\mathbf{x}), \ldots, f_f(\mathbf{x}) \right]
\quad \\ \text{s.t.} \quad \mathbf{x} \in \text{S}
\end{split}
\end{equation}

Where $x$ is the vector of the decision space that proposes a solution of minimizing $f$ and $f \in \mathbb{R}^m$. The notation 
Table \ref{tab:notation} illustrates all the notations in our research.

Our research employs the Non-dominated Sorting Genetic Algorithm III (NSGA-III)\cite{6600851} to address conflicted multi-objectives and propose solutions to minimise $f$. NSGA-III falls under the evolutionary multi-objective optimization (EMO) \cite{6600851}. The algorithm can find optimal solutions for up to 15 objectives, which is more than sufficient for our purpose.

\subsection{Load Balancing Model}

Early in this section, we model the data centre as a set of regions, where each region is composed of clusters. Each cluster is composed of PMs. Therefore, we assume the node to be either a bare metal or a VM with a fixed amount of resources and bandwidth. Meanwhile, we assume that the microservices are deployed to allocated containers that elastically consume the resources based on the requirements. In addition, a container is a placeholder; for example, $C_1$ hosts microservice $M_1$ therefore, we consider the container as the microservice $M_1$. The node can host a finite number of containers at the time $t$, thus, the total resource utilization of the node at the time $t$ is calculated as follows: 
\begin{equation}\label{EQ:1}
    \begin{split}
\text{total\_CPU\_util}(t) = \sum_{1}^{m} \text{CPU\_util}(M_m, t) \\
\text{total\_Memory\_util}(t) = \sum_{1}^{m} \text{Mem\_util}(M_m, t)
\end{split}
\end{equation}

It is well-established that microservices exhibit high CPU utilization with relatively lower memory utilization. This affects load balancing across resources at the node level. Therefore, a threshold for resource utilization is introduced to mitigate system performance issues resulting from the overloaded node. Thus, equation \ref{EQ:1} becomes: 
\begin{equation}\label{EQ:2}
    \begin{aligned}
\text{total\_CPU\_util}(t) = \sum_{1}^{m} \text{CPU\_util}(M_m, t) \\
\text{total\_Memory\_util}(t) = \sum_{1}^{m} \text{Mem\_util}(M_m, t)\\
\text{where} \\
\text{total\_CPU\_util}(t)\leq \text{Th} \ \ \text{and}\\
\text{total\_Memory\_util}(t) \leq \text{Th}
\end{aligned}
\end{equation}

Where $Th$ is the maximum utilization threshold. 

On the other hand, balancing the load at the cluster level requires distributing the load evenly across all the nodes in the cluster. Therefore, the mean $\mu$ and standard deviation (SD) are utilized to measure the load balancing among nodes in the cluster as follows: 

\begin{equation}\label{EQ:3}
    \begin{aligned}
        \text{SD}=  \sqrt{\frac{1}{N_{total}} \sum_{p=1}^{N_{total}} \left(\text{PM}_p - \text{mean}\right)^2}
    \end{aligned}
\end{equation}
\begin{equation}\label{EQ:4}
    \begin{aligned}
        \text{mean} = \frac{1}{N_{total}} \sum_{p=1}^{N_{total}} \text{PM}_p
    \end{aligned}
\end{equation}

where $N_{total}$ is the total number of PM in the cluster.

\subsection{QoS Model}

Quality of Service (QoS) parameters are crucial in cloud computing, including reliability, availability, and SLA violations. Cloud reliability is achieved by minimizing the risk of microservice failure. We, however, aim to reduce the risk of failure by scheduling microservices of one application across different PMs to prevent a single point of failure.

The availability of a microservice relies on its accessibility over the internet at any given time. Microservices may be offline during peak times when traffic is heavy, which is unacceptable. Additionally, network congestion leads to the blocking of services and degradation in availability. In our research, we aim to reduce network traffic costs to enhance microservice availability, as follows:

\begin{equation}\label{EQ:5}
    \begin{aligned}
    \textit{min \ Network \ traffic Cost}=\sum NT \\ 
    \end{aligned}
\end{equation}
where $NT$ is the network traffic in the cluster as communication between dependent microservices. 

SLA violations occur when the CPU of the host machine is utilized intensively. According to \cite{beloglazov2012optimal}, SLA violations could occur in two circumstances: during the migration of virtual machines (VMs), which in our study is the node (PM), or when the CPU of the host is utilized by 100\%. Migrating VMs degrades performance; hence, the Power Difference Model (PDM) proposed by \cite{beloglazov2012optimal} is utilized in the case of VM migration. The following equation represents the PDM:

\begin{equation}\label{EQ:6}
PDM=\frac{1}{M_{vm}}\sum_{j=1}^{M}\frac{C_{dj}}{C_{rj}}
\end{equation}

where $M_{vm}$ is the number of VMs, and $C_{dj}$ represents the estimation of system degradation caused by migrating VM$_j$.  $C_{rj}$ is the required amount of CPU by VM$_j$ during its lifetime.

In addition, we calculate the SLA violation utilizing the SLA violation time per active host (SLATAH) equation proposed by\cite{beloglazov2012optimal}: 
\begin{equation}\label{EQ:7}
    SLATAH = \frac{1}{N_{pm}}\sum_{i=1}^{N}\frac{T_{ui}}{T_{ai}}
 \end{equation}
 
where $N_{pm}$ is the total number of PMs in the cluster, $T_{ui}$ is the total time that $PM_i$ experiences 100\% CPU utilization, and $T_{ai}$ is the total number of active hosts. 

\subsection{Reliability Model}

Reliability refers to ensuring services are delivered and are available. In more precise terms, reliability measures the probability that scheduled services are available. According to \cite{bharany2022energy}, several circumstances can lead to service unavailability or failure, including 1) hardware failure, 2) operational failure, 3) network failure, and 4) software failure.

As these failures are negligible and irrelevant to the scheduling process, our research does not consider them. Cloud applications may be unavailable due to network conditions, preventing some clients from reaching the application. Scaling microservices by replicating them across different PMs becomes crucial to enhancing their availability. However, this poses challenges, as it can increase network traffic between microservices and complicate the architecture.

The goal of our research is to improve the availability of microservices by scheduling replicas in the most suitable PMs, thus enhancing availability and reducing network traffic. Microservices availability and the reduction of network traffic are conflicting objectives. Optimizing those two objectives is done by scheduling dependent microservices to the same PM. However, it increases the risk of a single point of failure. Conversely, scheduling dependent microservices across multiple PMs reduces the risk of failure but increases network traffic.

To evaluate the reliability of scheduling, entropy theory can be applied to measure the variance of microservices across PMs. By measuring entropy, we can assess the diversity and distribution of microservices on each PM, as well as clustering them, by using the following equation\cite{6773024}:

\begin{equation}\label{EQ:9}
    H = -\sum_{i=1}^{n} p_i \log(p_i)
\end{equation}
Where $ p_i$ is the probability of microservices scheduled into PM. 

To assess the quality of the scheduling, the SD and mean of entropy for the cluster are utilized as follows: 
\begin{equation}
    \begin{split} 
    \bar{H} = \frac{1}{N_{pm}} \sum_{j=1}^{n} H_j\\
    \sigma_H = \sqrt{\frac{1}{N_{pm}} \sum_{j=1}^{m} (H_j - \bar{H})^2}
    \end{split}
\end{equation}

Where $N_{pm}$ is the total number of PMs and $H_j$ is the entropy of $PM_j$.

\subsection{Latency Model}

Latency is a crucial aspect of QoS parameters promised by cloud providers. As microservices in certain applications scale due to client demand, latency becomes an issue. For instance, during peak traffic hitting application $A$ (e.g., an IoT application), the obvious solution is to create more replicas of the application tasks. The replicas in our case are microservice replicas. Each replica is considered a task that communicates with others. This leads to increased network traffic that may cause latency. Therefore, to minimize latency, we replicate microservices close to clients and close to each other. 

As reported by Alibaba data center, the latency $L$ between two different data centers located in Silicon Valley, USA, and Hangzhou, China, is measured to be 175 ms \cite{WebRef3}. We aim to utilize this metric to measure latency in our study as follows:

\begin{equation}\label{EQ:8}
    \begin{aligned}
        \text{Latency(t)}=\sum_1^l{D_{{pm},{pm+1}}}*l_v\\
        D_{{pm},{pm+1}}=\Biggl\{^{1 \ \text{if there are dependencies between {pm},{pm+1}}} _ {0  \ \text{otherwise}}
    \end{aligned}
\end{equation}

This equation calculates the latency at time $t$ by determining the dependencies between PMs $pm, pm+1$ that host two dependent microservices. It then multiplies that by the latency value $l_v$ considered to be 175 ms.

\subsection{Double Deep Q-network Model}

Reinforcement learning (RL) is one of the machine learning models that learn from interacting with the environment. RL finds the optimal policy by maximizing the expected cumulative reward over time \cite{chraibi2023novel}. The optimal policy is represented as a Q-function, which maps the state-action pair to the valued $Q$ as follow\cite{chraibi2023novel}s: 
\begin{equation}\label{EQ:10}
    Q \colon S \times A \rightarrow \mathbb{R}
\end{equation}

Where $S$, and $A$ are state and action values, respectively. 

The Q-function is estimated using various algorithms\cite{chraibi2023novel}. The Q-value represents the action value of a specific state $s_t$ taking action $a_t$ at time $t$. The Bellman equation is employed to estimate the Q-value in the following equation\cite{chraibi2023novel}: 

\begin{equation}\label{EQ:11}
\small
Q(s_t, a_t) = (1 - \alpha) \cdot Q(s_t, a_t) + \alpha \cdot \left( R_{t+1} + \gamma \cdot \max_{a_{t+1}} Q(s_{t+1}, a_{t+1}) \right)
\end{equation}

Where $(1 - \alpha) \cdot Q(s_t, a_t)$ is the current Q-value. $\alpha R_{t+1}$ is the reward obtained from current action $a_t$ at state $s_t$, and $\gamma$ is the discount rate associated with reward to give a more important recent reward. $\max_{a_{t+1}} Q(s_{t+1}, a_{t+1})$ is the maximum reward that can be obtained in state $s_{t+1}$

The Q-function can handle a small number of states \cite{chraibi2023novel}\cite{wang2022dueling}. Therefore, the deep Q-networks are utilized in our research. A neural network can approximate the action-value function as follows\cite{chraibi2023novel}: 

\begin{equation}\label{EQ:12}
Q^*(s,a) = \mathbb{E}_{s_{t+1} \sim E} [r + \gamma \max_a Q^*(s_{t+1},a_{t+1})]
\end{equation}

The intuition behind this equation is as follows: if the optimal value of the next step $Q^*(s_{t+1},a_{t+1})$ is known, then the optimal next action $a_t$ is selected to maximize the reward.

Non-linear function approximators, like neural networks, are utilized in our research. A q-network with weights $\theta$ is used to approximate the action-value function. It maps the state of the approximate function using wight $\theta$ as proposed in the equation (\ref{EQ:12}) \cite{mnih2013playing}. In addition, the loss function proposed in \cite{mnih2013playing} is utilized in our research as follows: 

\begin{align}\label{EQ:13}
\small
\nabla_{\theta_i} L_i (\theta_i) = \mathbb{E}_{s,a \sim \rho(\cdot); s \sim \mathcal{E}} 
&\left[ r + \gamma \max_a Q(s,a;\theta_{i-1}) \right. \nonumber \\
&\left. - Q(s,a;\theta_i) \nabla_{\theta_i} Q(s,a;\theta_i) \right]
\end{align}

In each iteration, the $\theta$ is updated utilizing the Bellman equation into the target Q-value.

Our agent is model-free, adapting to the dynamic environment. It learns directly from interactions by exploring different actions and observing resulting states to maximize rewards. The agent is off-policy, employing a $\epsilon$-greedy policy to select actions $a$ that maximize the Q-function. We use an $\epsilon$-greedy strategy to balance between exploitation and exploration. Where $\epsilon$ selects the optimal actions, and $1-\epsilon$ selects random actions. Figure \ref{fig:The Agent Architecture} illustrates the agent components while training. 

\section{Framework Architecture}

\begin{figure*}[h!]
     \centering
     \begin{subfigure}[b]{0.50\textwidth}
         \centering
         \includegraphics[width=\textwidth]{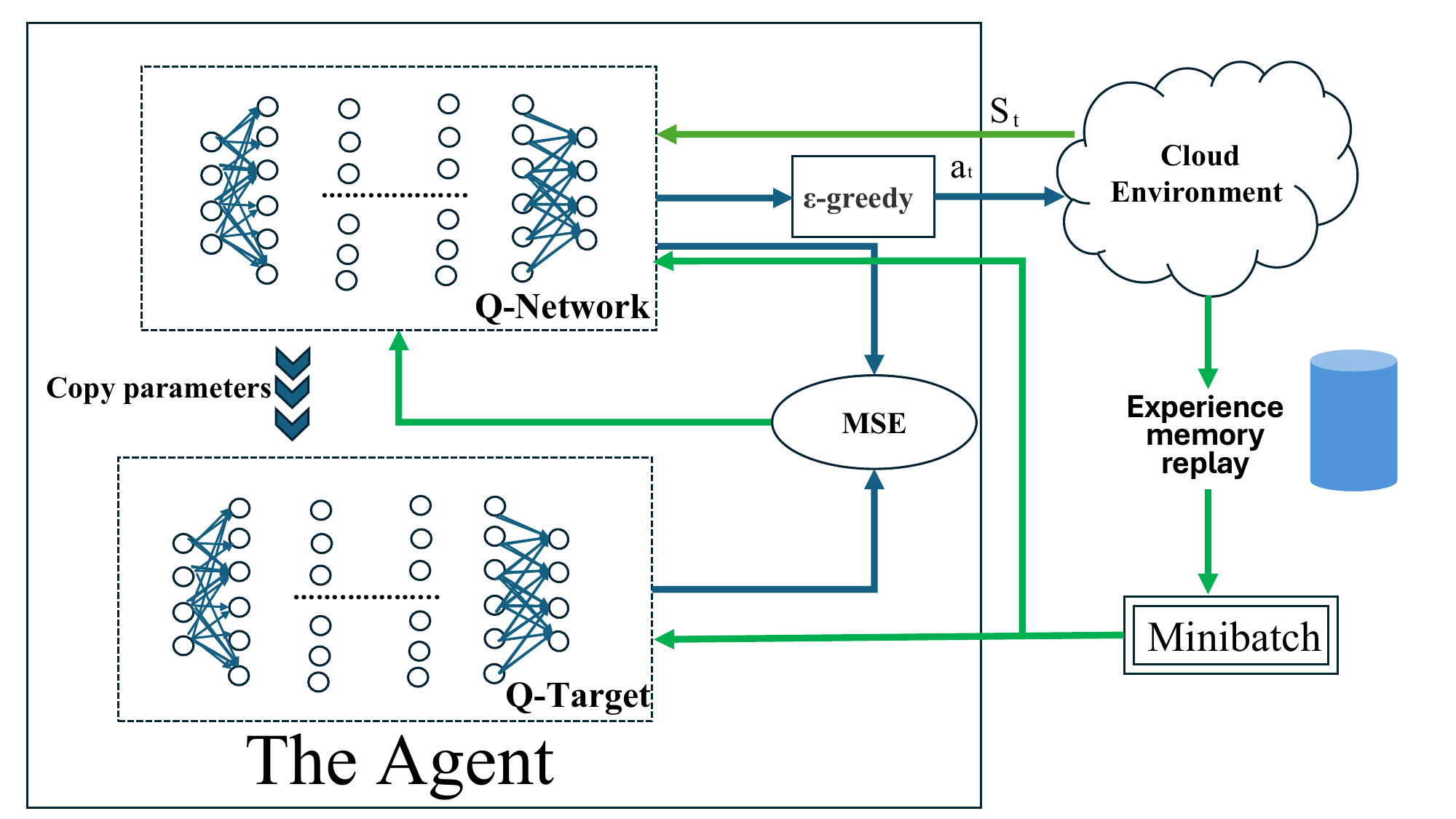}
         \caption{The Agent Architecture}
         \label{fig:The Agent Architecture}
     \end{subfigure}
     \hfill
     \begin{subfigure}[b]{0.49\textwidth}
         \centering
         \includegraphics[width=\textwidth]{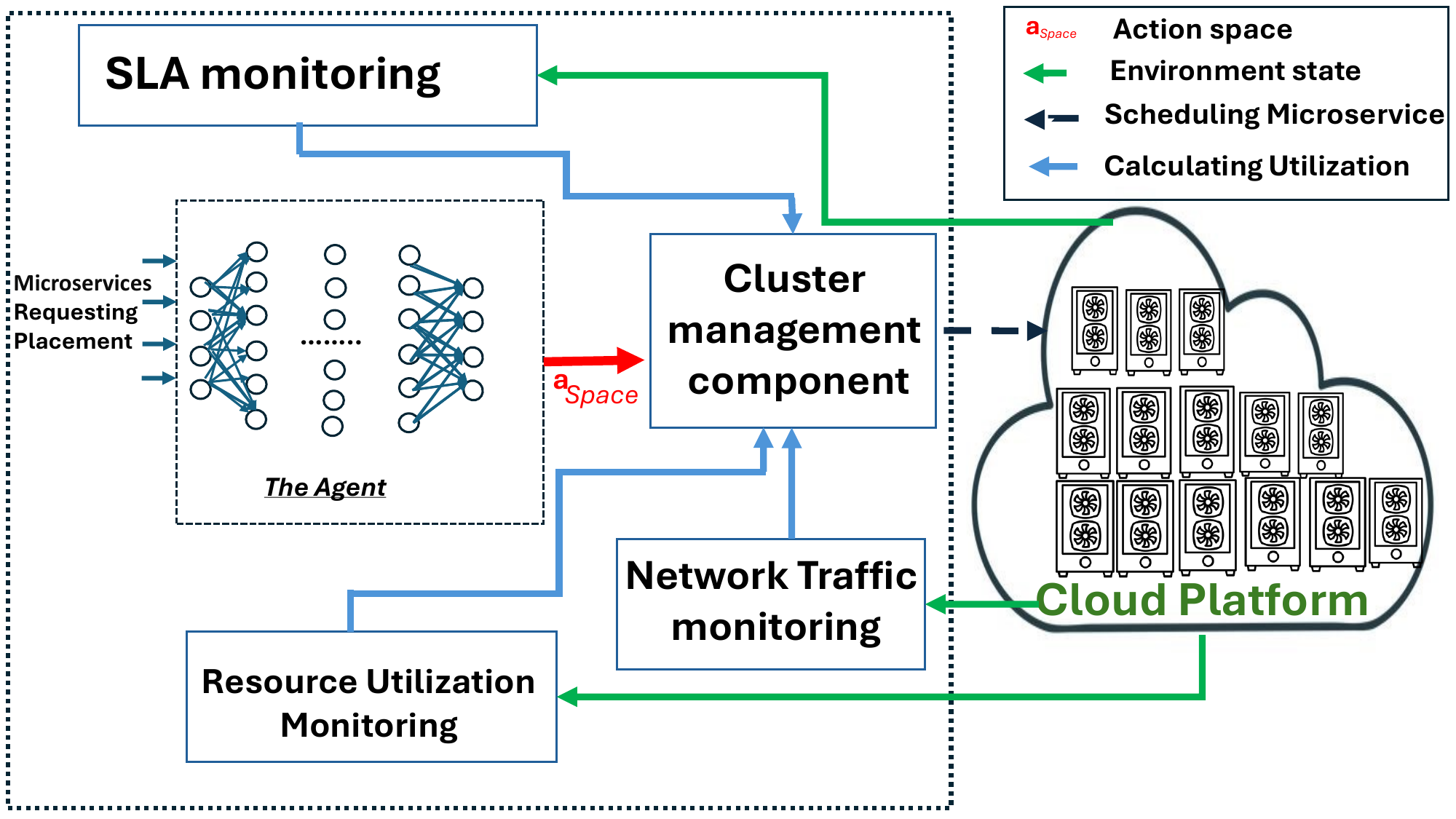}
         \caption{The DQN-Scheduler Architecture}
         \label{fig:The Framework Architecture}
     \end{subfigure}
     \caption{The figure shows the agent training steps alongside our proposed scheduling framework.}
     \end{figure*}

In this section, we describe the components of DQN-Scheduler. The components handle scheduling microservices into cloud computing, aiming to enhance the total performance of the cloud by enhancing resource utilization, reducing network traffic, mitigating the risk of violating SLAs, and reducing latency. DQN-Scheduler consists of five components, as illustrated in Figure \ref{fig:The Framework Architecture}.
% \begin{figure}[ht!]
%     \centering
%     \includegraphics[width=0.55\textwidth]{the_framework.pdf}
%     \caption{The Framework Architecture}
%     \label{fig:enter-label}
% \end{figure}
\subsection{The Agent} Here, we briefly outline the agent in two phases: implementation and training. In the first phase, the role of the agent in scheduling is described. In the second phase, we explain the training process of the model.
\subsubsection{The Implementation Phase}
The agent is used to provide the DQN-Scheduler with scheduling decisions. The scheduling decision space is represented as a vector of the probability of each PM in the cluster hosting incoming microservices. 

\subsubsection{The Training Phase} 

The architecture of the agent consists of \textit{state}, \textit{action}, \textit{environment}, and \textit{reward} as shown in Figure \ref{fig:The Agent Architecture}. 

\textit{The state:} is an input consisting of the resource requirements of incoming jobs (microservices). Typically, clients submit requests to schedule their jobs with requirements. The requirements vary depending on each client's application. Therefore, instead of solely using the client's resource requirements, we consider the maximum, minimum, and average resource requirements during model training. Thus, the \textit{state} is denoted as 
\begin{equation}
\begin{split}
S = \Bigl[ 
[max(s_j^{CPU}), avg(s_j^{CPU}), min(s_j^{CPU})], \\
[max(s_j^{Mem}), avg(s_j^{Mem}), min(s_j^{Mem})]
\Bigr]
\end{split}
\end{equation}

where $j$ denotes the job coming to the cloud to be scheduled, and \textit{max}, \textit{avg}, and \textit{min} represent the maximum, average, and minimum resource requirements for CPU and memory, respectively. Additionally, the state includes the number of repliacs specified by the client, denoted as $n_{replica}$.

The agent comprises two neural networks: Q-network and Q-target. Both networks share the same architecture, consisting of two convolutional layers (CNN) followed by two fully connected layers (linear layers), and a classifier layer. The ReLU() activation function is employed in both networks to leverage non-linearity in learning the optimum Q-function. The ReLU activation function is specifically utilized to mitigate the issue of vanishing gradients. Additionally, Softmax() is utilized to convert logits to action probabilities. 

The loss, computed as shown in equation (\ref{EQ:13}), leads to unstable weight updates in the Q-network. Therefore, periodic weight copying between the Q-network and Q-target networks is performed as shown in Figure \ref{fig:The Agent Architecture}. 
Furthermore, experience memory replay is employed to store recent state-action and next-state events for training the Q-target network. 

\textit{The action:} is a component that determines microservice scheduling into PM, with the agent output representing the action space. This action space provides an action vector, denoted as $a=\{a_1,a_2...a_a\}$. Each action in the vector represents a probability that a particular PM will host the microservice. Softmax function is utilized to select the maximum action probability that maximizes the reward. In addition, $\epsilon$-greedy exploration is employed to explore other potential actions that may maximize reward using random action selection.

\textit{Reward} is utilized to encourage the agent to behave according to the aims of our research, which are to enhance resource utilization, and latency, balance the load, and manage QoS parameters. These objectives are conflicting. Therefore, NSGA-III is utilized to find the optimum solution among those objectives using equation (\ref{EQ:9}). Calculation of the reward is based on the distance between the optimal solution of all objectives given by NSGA-III and the current state of the environment, as follows:

\begin{equation}
\label{EQ:14}
\text{ED} = \sqrt{\sum_{i=1}^{i} ( \text{Pareto\_Front}_i -\text{objective}_v )^2} 
\end{equation} 

$ED$ represents the Euclidean distance between the current state and each point in the Pareto front. The $objective_v$ are the differences between the current environment and the environment after scheduling the microservice, such as $\Delta$ CPU utilization, $\Delta$ memory utilization, and $\Delta$ network traffic. Some of these differences may be negative, such as network traffic. Therefore, calculating $ED$  becomes: 

\begin{equation}\label{EQ:15} 
\text{ED} = \sqrt{\sum_{i=1}^{v}(\text{Pareto\_Front}_i - \begin{bmatrix} \Delta\text{CPU\_uti} \\ \Delta\text{Mem\_uti} \\ \Delta\text{network\_traffic} \end{bmatrix})^2} \end{equation}

\subsection{SLA Monitoring}

This component of the DQN-Scheduler is responsible for reporting any service violations occurring during scheduling, like rejecting scheduled microservices. It also reports any violation that occurs after scheduling, like being unable to reach the service due to network congestion.

During the time of training the model, SLA monitoring, resource monitoring, network traffic monitoring, cloud cluster management, and resource monitoring were parts of the environment. They also contribute to reward calculation. For example, if the model violates the SLA or schedules a microservice to a PM exceeding the threshold, the agent is rewarded with negative values. 

SLA monitoring components aim to keep an eye on resource utilization and enforce the threshold restrictions all the time by applying equations (\ref{EQ:1}) and (\ref{EQ:2}). The aim is also to report any service violation during the migration process or if the CPU is being utilized intensively, using (\ref{EQ:6}) and (\ref{EQ:7}).

\subsection{Resource Utilization Monitoring}
This component of the DQN-Scheduler measures resource utilization at timestamp $sp$, where the total proportional resource utilization is calculated as follows:

\begin{equation}
\begin{split}
   \ \ R_{cpu} =\frac{total \ CPU \ util(sp)}{C_{cpu}}, and \\  R_{mem} =\frac{{total \ memory \ util(sp)}}{C_{mem}}
  \end{split}
\end{equation}

Here, $total \ CPU \ util(sp)$ represents the total CPU utilization at timestamp $sp$ provided by equation (\ref{EQ:3}). $C_{cpu}$ denotes the total CPU capacity in a PM. Similarly, $R_{Mem}$ calculates the memory utilization at timestamp $sp$ for the PM.

This component is responsible for reporting the resource utilization of the PMs at timestamp $sp_p$. Additionally, the component calculates the expected resource utilization when an incoming microservice is scheduled in real-time for both resources (CPU and memory), denoted as $Exp^n_{cpu}$ and $Exp^n_{mem}$, respectively, as follows:
\begin{equation}
\begin{split}
    Exp^n_{cpu} = \frac{Cur\_R_{cpu}^n + Req\_R_{cpu}^n}{C^n_{cpu}}, and \\
    Exp^n_{mem} = \frac{Cur\_R_{mem}^n + Req\_R_{mem}^n}{C^n_{mem}}     
\end{split}
\end{equation}

Where $Cur\_R$ and $Req\_R$ represent the current utilization of the resource and the microservice request of the resource, respectively.

\subsection{Network Traffic Monitoring}
Three types of network traffic are considered in microservice architecture: traffic between microservices located in the same zone, network traffic originating from clients to access distributed services, and traffic between microservices in different zones. The first two types of network traffic result in negligible latency due to the high speed of the network (e.g., 5G). In addition, they are managed by cluster orchestration (in our case, Kubernetes). However, network traffic monitoring calculates the latency caused by the third type using Equation (\ref{EQ:8}).

This component reports network traffic latency to the cluster management component during microservice scheduling, including the increases or decreases in network latency. If the cluster management component identifies that latency would lead to SLA violations or performance degradation, microservice scheduling is reassessed. During the training of the agent, we utilized this component by associating a negative value with the reward each time latency increased. This step aims to encourage our agent to reduce latency. Additionally, the component notifies the DQN-Scheduler of the changes in latency each time a microservice is scheduled to maintain QoS and prevent network latency or congestion Figure\ref{fig:The Framework Architecture}.

\subsection{Cluster Management Component}

The component coordinates between the agent and cloud resources (Figure \ref{fig:The Framework Architecture}) and fulfils several key roles:
\begin{enumerate}
 \item Receives resource allocation orders for deployed containers from the agent.
 \item Collects the current cluster status, including resource utilization, latency, and load balancing. 

\item Assesses the agent's decisions regarding microservice deployment locations. 

\item Deploys and runs microservices within deployed containers. \end{enumerate}

The agent provides decisions of scheduling microservices as a decision space instead of a single decision. For instance, instead of the agent designating a single PM qualified to host the microservice, probabilities for all PMs hosting a microservice are provided. After this component gathers the necessary data from both the agent and environment, the scheduling process commences. If the best decision to schedule microservices violates the SLA or degrades system performance, decisions from the action space are updated using the Bayesian inference method. 

A posterior probabilities method is used to update the probabilities of hosting microservices by PM based on new evidence. In our case, the evidence is the probability of QoS parameters and it is calculated as the following equation as it is presented in \cite{ellison2004bayesian}:

\begin{equation}
y = \frac{1}{1 + e^{-\log{\left(\frac{x}{1-x}\right)} + \log{\left(P(R|n)\right)}}}
\end{equation}

Where $\log{\left(P(R|n)\right)}$ represents the likelihoods. The likelihood is either 0 or 1, determined based on the evidence, threshold, and microservice request for resources. Additionally, $\frac{1}{1 + e^{-\log{\left(\frac{x}{1-x}\right)}}}$ represents the prior probabilities calculated using the logistic function. 

The Bayesian inference model does not make decisions but reorders them based on the current environmental status. This reordering involves updating the probability of hosting microservices for all PMs. The aim is to reduce SLA violations and performance degradation. Therefore, this component is utilized during agent training and microservice scheduling.

\begin{table}[]
\centering
\caption{Table of Notation}
\begin{tabular}{|p{2cm}|p{6cm}|}
\hline
\textbf{Notation} & \textbf{Description} \\
\hline
$D$ & Data center consisting of a set of data zones. \\
$C$ & Cloud cluster consisting of a set of nodes. \\
$P$ & Set of pods allocated to a node. \\
$R$ & Set of containers in a pod. \\
$M$ & Microservice. \\
$M_{replica}$ & The number of replicas of a microservice. \\
$O_d$ & Number of objectives in multi-objective optimization. \\
$z_1, z_2, \ldots, z_z$ & Data zones within a data center. \\

$n_1, n_2, \ldots, n_n$ & Nodes within a cloud cluster. \\

$p_1, p_2, \ldots, p_p$ & Pods within a node. \\

$r_1, r_2, \ldots, r_r$ & Containers within a pod. \\
$S$ & Feasible solution space. \\
$\mathbf{x}$ & Decision variable in the optimization problem. \\
$PM$ & Physical machines. \\
$VM$ & Virtual machines. \\
$\text{Th}$ & Maximum resource utilization threshold. \\
$SD$ & Standard deviation. \\
$\mu$ & The Mean. \\
$H$ & Entropy, used to measure the diversity of microservices across physical machines. \\
$L$ & Latency between data centers or nodes. \\
$Q(s, a)$ & Q-function, representing the state-action value in reinforcement learning. \\
$\alpha, \gamma$ & Learning rate and discount rate in reinforcement learning. \\
$Q^*(s, a)$ & Optimal Q-value in reinforcement learning. \\

$R_{cpu}$ & Proportional CPU utilization. \\
$R_{mem}$ & Proportional memory utilization. \\
$Exp^n_{cpu}$ & Expected CPU utilization in $PM_n$. \\
$Exp^n_{mem}$ & Expected memory utilization in $PM_n$. \\
$Cur\_R_{cpu}$ & Current CPU utilization. \\
$Req\_R_{cpu}$ & Requested CPU utilization by a microservice. \\
$L$ & Network latency between zones. \\
$H$ & Entropy, measuring diversity across physical machines. \\
$Q(s, a)$ & Q-function, representing state-action value in reinforcement learning. \\
$y$ & Logistic function used in Bayesian inference for scheduling. \\
$P(R|m)$ & Likelihoods for hosting microservices, based on resource utilization. \\
\hline
\end{tabular}
\label{tab:notation}
\end{table}

\subsection{The Algorithm Pseudocode}
In this section, the algorithm for scheduling microservices into the cloud is explained. The algorithm describes the collaboration among different components of DQN-Scheduler and illustrates the sequential flow of operations.
\begin{algorithm}
\caption{Scheduling Framework Using Q-network Agent()}\label{alg:1}
\begin{algorithmic}[1]
    \State Input: Node List $N$,  client microservices m($rec_{cpu}$,  $rec_{mem}$)
    \State Output: R $\leftarrow$ Inst \Comment{allocating R} 
    \State Output: Inst $\leftarrow$ m \Comment{Scheduling}
        
    %\For{each inst $\leftarrow$ instances }
        \For{each m $\leftarrow$ microservices }
        \Procedure {collecting Evedince}{$N,m$}%\Comment{Current state of env}
        \State Set $ rec_{cpu} \leftarrow R_{cpu}$
        \State Set $ rec_{mem} \leftarrow R_{mem}$
        \State Set $ l \leftarrow L$ \Comment{latency}
        \State Set $ LB \leftarrow SD$ \Comment{load balancing}
        \State Set $ SoQ \leftarrow SLATAH()$ \Comment{SLA violation}
        \State return $s \leftarrow \left[R,l,LB,SoQ\right]$
        \EndProcedure
        \Procedure{PredictDecisionSpace}{$m$}
        \State Initialize Q-network agent $Q$
        \State Set $D \leftarrow Q(m)$ %\Comment{Predict decision space using Q-network}
        \State \textbf{return} $D$
        \EndProcedure
        %\State D $\leftarrow$ target Q-network\Comment{D is decisions space}
         \Procedure {cluster\_Managmrent}
         {$D,s$}
            \State Set $D^ \prime \leftarrow posterior\_probabilities ()$
             \EndProcedure
             \State return $D^ \prime$
            \State Set $D^ \prime \leftarrow argmax(D^ \prime,n_{replica})$
            \Procedure {Scheduling}{$D \prime, rec_{cpu},rec_{mem}$}
            \State Setup scheduling configuration
            \State  R  $\leftarrow allocatResource(Inst)$
            \State  Inst $\leftarrow scheduling(m)$
            \State Set $Scheduling\_flag \leftarrow True$
            \State return  $Scheduling\_flag$
            \EndProcedure
            \If {$Scheduling\_flag$}
            \State break
            \EndIf
    \EndFor
       %  \EndFor
\end{algorithmic}
\end{algorithm}

Algorithm \ref{alg:1} states that $N,m$ are taken as inputs for the cloud environment. $N,m$ includes cloud cluster PMs and resource requirements by microservices, respectively. The expected output of the DQN-Scheduler is a scheduled microservice allocated to a resource, identified as an instance denoted by $Inst$.

From lines 5 to 11, the algorithm collects the status (evidence) of the current resource utilization of PMs in the cluster. Lines 13 to 16 illustrate the process of utilizing the Q-network agent for decision-making. At this stage, our agent does not require the environment status. Therefore, it predicts scheduling decisions based on the provided state. The scheduling decision is formed as a vector of probabilities, indicating the probabilities of each PM hosting the microservice $m$.

We aim to mimic the stochastic nature of cloud computing environments. In addition, we expect an instability in the rate of scheduling microservices in the cloud. Thus, lines 18 and 19 demonstrate the use of Bayesian inference to update the decision space if necessary.

The remaining lines of the algorithm (lines 21 to 33) select the optimal PMs from the decision space to host the microservice. The last step is to deploy the microservice to the selected PM. 

Using Bayesian inference allows the DQN scheduler to adapt to rapid changes in the cloud environment, enhancing the reliability and efficiency of microservice scheduling. Updating the decision space continuously allows the DQN-Scheduler to optimize microservice deployment in real-time.

\section{Scenario and Cloud Configurations}

The characteristics of the cloud environment and the tasks are briefly discussed in this section. The aim is to provide an overview of the assumptions regarding the scenario and cloud configurations.

\subsection{Workload Characterization}

\textit{Batch tasks}: These are short-running applications deployed directly to the PM. Each application is usually divided into several tasks that are deployed separately on the instance, providing enough resources. They represent just 10\% of applications submitted to the cloud \cite{guo2019limits}\cite{zhong2020cost}\cite{9242282}.

\textit{Online services}: These are applications that run for an extended period and represent the majority of deployment to the cloud\cite{guo2019limits}\cite{zhong2020cost}\cite{9242282}. They are considered production applications consisting of user-facing services such as search engines, e-commerce, and online shopping applications. They can be scaled across cloud PMs to provide sufficient access to the service, maintaining availability.

The two types of applications are typically scheduled through a centralized approach in well-known data centres, such as Brog \cite{verma2015large}\cite{reiss2011google} and Quasar\cite{delimitrou2014quasar}. Alibaba data centre utilizes Fuxi for scheduling batch jobs \cite{zhang2014fuxi}, whereas Sigma is used for long-running applications \cite{AlibabDataTrace2}.

We consider the two types of workload, both of which exhibit dependency. For instance, the batch workload is divided into tasks that rely on each other and are scheduled within microservices. Similarly, online services consist of multiple services communicating with each other. Therefore, both types are categorized in our study as dependent microservices.  

\subsection{Stochastic Cloud Environment} 
In our research, a stochastic cloud environment is considered to mimic the reality of scheduling microservices. Several factors are taken into account to simulate the scheduling scenario.

\textbf{\underline{Timing:}} We assume that the number of scheduled tasks varies over time. Applications with multiple tasks arrive in the scheduling queue at different timestamps. Tasks are then scheduled individually while maintaining dependencies between them. A stochastic environment is simulated, with various job types continuously and periodically submitted to the cloud across multiple clusters. 
 
\textbf{\underline{Task Priorities:}} Online service tasks take priority over batch tasks. Tasks failing to execute or requiring rescheduling are placed at the end of the queue without priority escalation.

\textbf{\underline{Scaling Applications:}} In microservices architectures, scaling applications is common and can be achieved through two methods: \begin{enumerate}  
\item  Submitting scheduling requests with specified task replicas.
\item Replicating tasks of running applications to manage outbound traffic.
\end{enumerate}

\textbf{\underline{Resource Utilization:}} We assume that different types of microservices utilize heterogeneous resources in real-time.

\textbf{\underline{Cloud Zones Network Traffic:}} A cloud data center consists of graphical zones linked by high-speed networks. PMs within the same zone experience manageable network latency, which we do not focus on in this research. However, we consider the costly network traffic between zones.

\section{Baseline algorithms}

In this section, the baseline algorithms used to compare with the DQN-Scheduler are explained briefly. Those algorithms are well-known and have been utilized in many fields.

\subsection{Genetic Algorithm (GA):} GA is a population-based meta-heuristic algorithm that provides multiple solutions\cite{katoch2021review}\cite{michalewicz1999genetic}. The GA optimization algorithm is inspired by the natural selection mechanism. It produces a diverse population of solutions that prevent getting stuck in local optima \cite{katoch2021review}.  It starts with the population of chromosomes representing the solution needed. 

Three biological operations are used to find the optimum chromosomes, including selection, crossover, mutation, and fitness function. The chromosomes represent the microservice placement into the PMs. The chromosomes are initially generated randomly. The chromosomes are chosen based on a fitness function. The fitness function of our research aims to enhance resource utilization and reduce network traffic consumption. Crossover and mutation operations are utilized to explore the solution space.

\subsection{Particle Swarm Optimization (PSO):} PSO is a meta-heuristic algorithm that explores the solution space for optimal solutions. It mimics natural species like bees that navigate in space to find their destination\cite{kalimuthu2024design}. Like GA, PSO utilizes a fitness function to assess the fitness of solutions. PSO consists of a population, velocity, and target destination parameters guiding particles towards their destinations. The population is generated based on the resource requirements of microservices. In addition, velocity guides particles as a group towards optimal positions. It is influenced by the positions of both the particle and the group \cite{alelyani2024optimizing}.

We conceptualize microservice resource requirements as particles. Whereas PMs represent destinations spread across the solution space, offering CPU and memory resources for the particles. Once particles explore the space, a fitness function evaluates each particle solution's fitness. The fitness function aims to enhance resource utilization and reduce the cost of utilizing network traffic. 

\subsection{Best Fit Algorithm (BFA):}
The best-fit algorithm is employed to tackle the bin packing problem (BPP). BPP is recognized as an NP-hard problem. Finding the optimal solution for BPP within polynomial time is impossible. BFA serves as a heuristic like first-fit, worst-fit, and next-fit algorithms \cite{kaaouache2015solving}. 

The BFA is utilized in our research to optimize the packing of microservices onto PMs while considering resource utilization, network traffic, and load balance. Therefore, we represent the resource requirements as the value of items needing packing. Whereas, we represent the available resources in each PM as the values of the bins. Then BFA is utilized to schedule microservices into PMs.

Upper and lower thresholds for resource utilization are established. When PM reaches the maximum threshold, it is removed from the list. Once PM's resource availability falls within the lower and upper thresholds, it is queued again. This threshold management ensures improved resource utilization. BFA clusters dependent microservices near each other during packing.

\section{Experiment Setup and Results }
This section outlines the experiment. The section begins with a description of the experiment setup. Second, the dataset used in the experiment and the results of training the agent utilized in the scheduling process are explained. The last part of this section details the experimental results.  
\begin{figure*}[h!]
    \centering
    \includegraphics[width=0.95\textwidth]{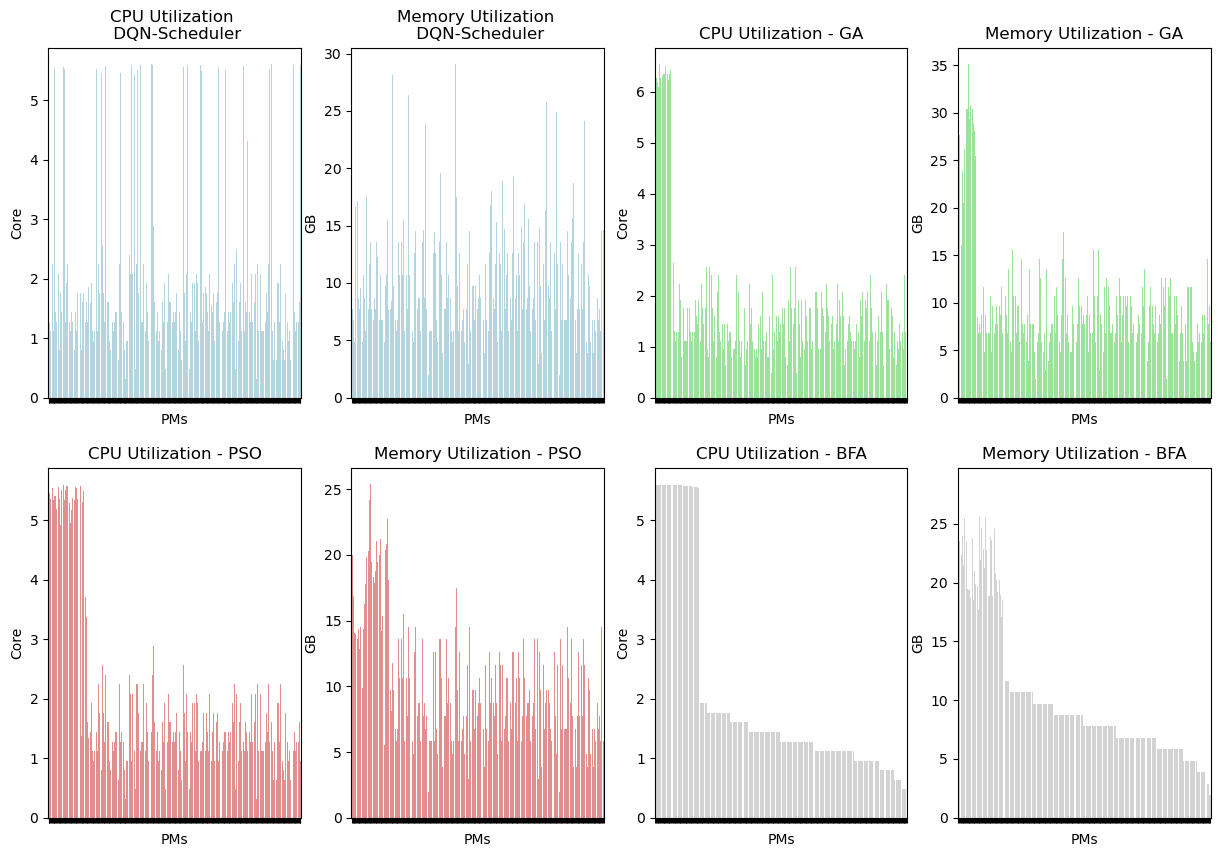}
    \caption{Resource Utilization by DQN-Scheduler and Benchmark Algorithms for Each PM Resource (CPU and Memory)}
    \label{fig:reource_utilization}
\end{figure*}

\begin{table*}[h!]
    \centering

    \caption{Performance Comparison: DQN-Scheduler and Benchmark Algorithms}
    \label{tab:algorithm_performance}
    \begin{tabular}{ccccccccccc}
\hline
& \multicolumn{2}{c}{CPU Utilization} & \multicolumn{2}{c}{Memory Utilization} & \multicolumn{2}{c}{QoS} & \multicolumn{2}{c}{Reliability}&\multicolumn{2}{c}{Latency} \\
        \cmidrule(lr){2-3} \cmidrule(lr){4-5} \cmidrule(lr){6-7}\cmidrule(lr){8-9} \cmidrule(lr){10-11}% Adding midrule for clarity
        Name & {SD}  & {Mean} & {SD}  & {Mean} & SLATAH & &{SD}  & {Mean} &{SD}  & {Mean}  \\
        \midrule
        GA & \textbf{0.73099} & 1.299 & 4.3192 & 9.073&0.000093&&0.9242&0.2605& 36.0443 & 9.0028 \\
        PSO & 1.00558 & 1.4432 & \textbf{3.2842} & 8.60123 &0&&0.4615&0.1412& \textbf{8.8421} & 3.459 \\
        BFA & 0.9398 & 1.3299 & 3.6014 & 8.4343 & 0&&0.4490&4.31443&15.8433 & 4.8918 \\
        \textbf{DQN-Scheduler} & \textbf{\textit{0.7499}} & 1.3043 & \textbf{\textit{3.3464}} & 8.57927 & \textbf{\textit{0}}&&\textbf{\textit{1.1033}}&0.4547&\textbf{\textit{12.7412}} & 4.8555 \\
        \bottomrule
    \end{tabular}
\end{table*}

\subsection{Experiment Setup}
The proposed framework was evaluated by simulating a cloud data centre. The simulation includes the workload of microservices, workload arrival times, data center capacity, network architecture, and container allocation for hosting microservices. An Alibaba dataset of the workloads of microservices was used. The cloud configuration was simulated based on real Alibaba resource specifications.

The experiment was conducted using Python 3 and PyTorch 121.coda. A computer featuring a 12$^{th}$ Gen Intel(R) Core(TM) i7-12650H processor running at 2.30 GHz was utilized. The machine used for the simulation is equipped with a GeForce RTX 4060 GPU with 16 GB of RAM. This machine was also used for training the agent.

\subsection{Dataset}
\begin{figure}[h!]
    \centering
    \includegraphics[width=0.45\textwidth]{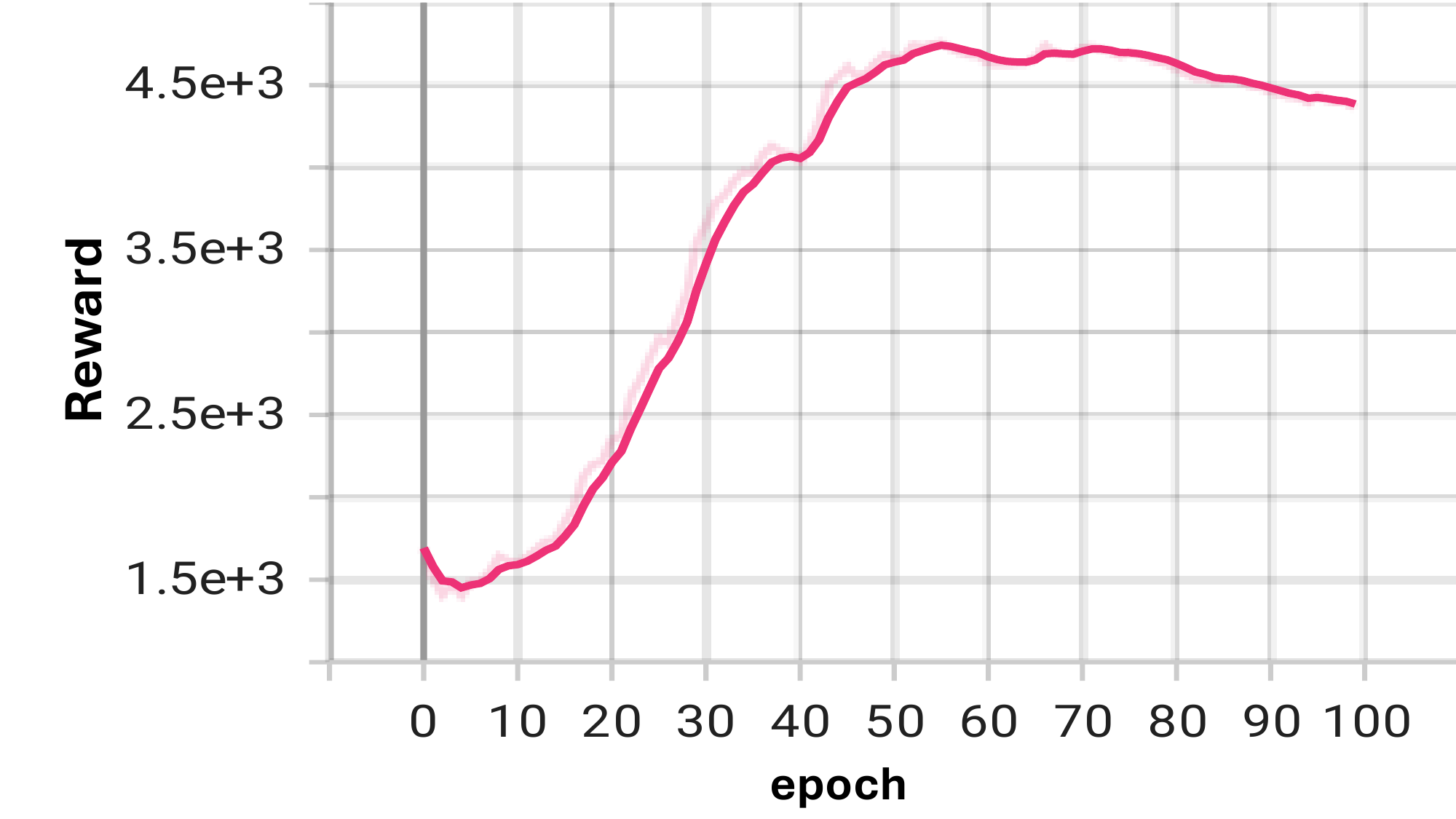}
    \caption{The Reward Scoring of the Agent During Training}
    \label{fig:TheResultOfTrainingTheAgent}
\end{figure}

Three types of datasets were utilized in our research, as follows:

\begin{enumerate}
\item Alibaba\cite{AlibabDataTrace}: This dataset is publicly available from the Alibaba data center. It contains essential details such as node IDs, timestamps, and comprehensive resource utilization information. It describes the infrastructure as a microservice-based data center. It details the resource utilization data of over 90,000 containers running microservices. The dataset of containers that are allocated to more than 1,300 PMs is included. The dataset consists of node ID, timestamp, and PM resource utilization details.

\item   A dataset describing the dependencies between microservices in the Alibaba data center, made public by \cite{ luo2021characterizing} was used. This dataset presents the communication between microservices as a direct call graph. Microservices are categorized into upstream microservices (UM) and downstream microservices (DM). In addition, the pattern of communication between microservices is described in the dataset. We extracted data from over 3,280 microservices running for one hour on 300 PMs.

\item Data centre configurations were synthesized, including zone names and capacity. The data seed for distributing stateful services to cloud resources was also synthesized. This step aims to simulate real-world data centers.

\subsection{ Agent Training Result}
In this section, the result of training the agent is briefly described. The aim is to show that the agent learns the scheduling process. The training results show that the agent is rewarded based on multiple objectives Figure \ref{fig:TheResultOfTrainingTheAgent}. The cumulative rewards gained by the agent indicate that the model is progressing in its learning. This also demonstrates that the agent has learned to schedule microservices with optimal solutions.

The reward scoring utilizes a decaying $\epsilon$-greedy algorithm. Initially, the agent starts with high randomness in its actions to explore the action space, gradually decreasing randomness as it learns to select more optimal actions. Figure \ref{fig:TheResultOfTrainingTheAgent} shows that the agent started to learn and selected close to optimal actions. The results demonstrate that the model maintains a consistent average reward, indicating that it selects nearly the optimal solution each time.

\end{enumerate}

\subsection{Discussing The Results}
This section details the experimental results, discusses system performance, and compares the DQN-Scheduler with benchmark algorithms.

\subsubsection{Load Balancing of Resource Utilization}
This subsection focuses on two key features of the DQN-Scheduler: enhancing resource utilization and balancing load across all PMs in the cluster. This section demonstrates the efficiency of the DQN-Scheduler compared to other approaches in these two aspects.

\begin{figure}[h!]
    \centering
    \includegraphics[width=0.50\textwidth]{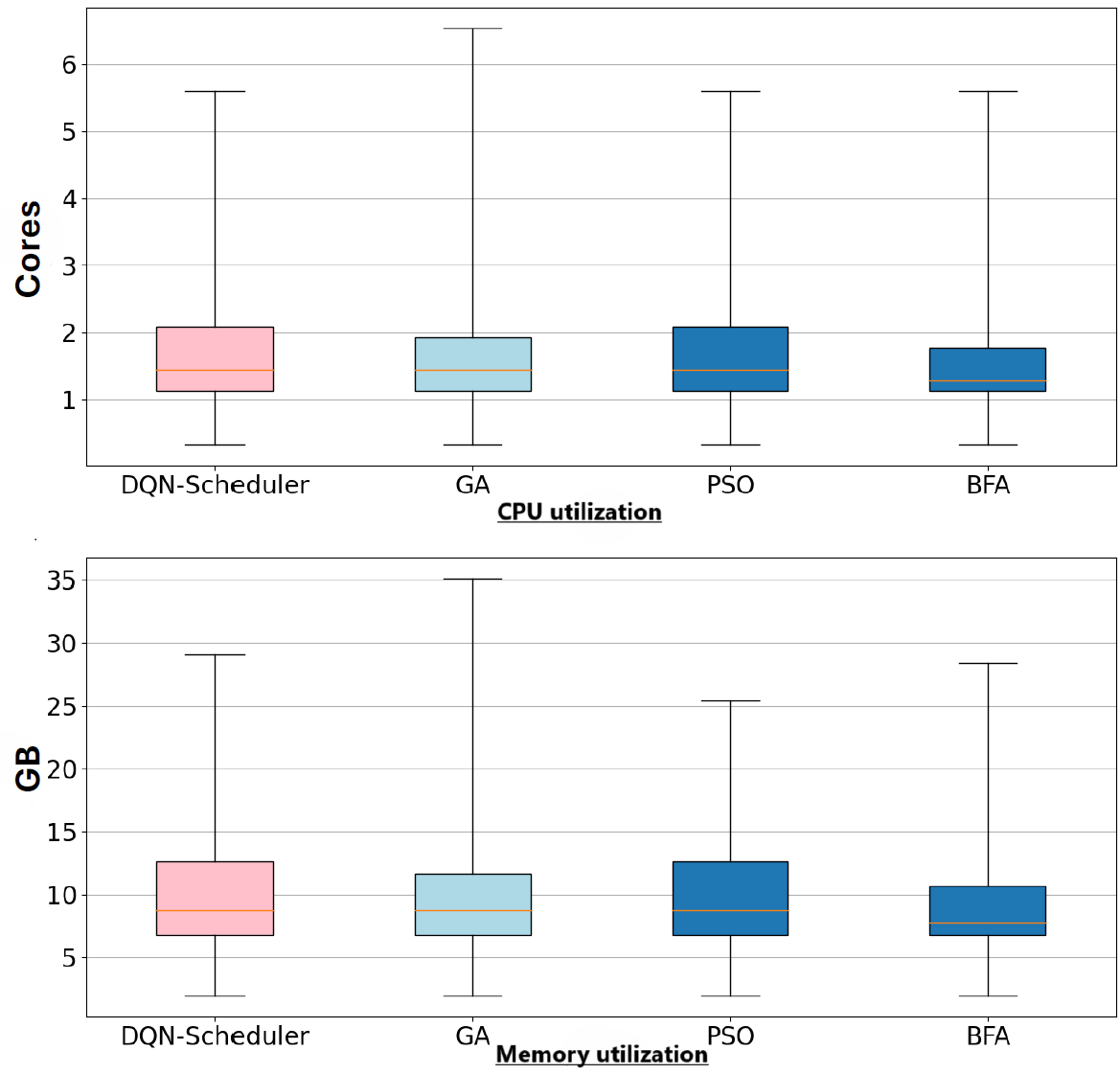}
    \caption{The Results of Balancing The Resource Utilization For Each Approach Utilizing: Standard Deviation and Mean}
    \label{fig:CPUandMemorySDandMean}
\end{figure}

\textbf{Enhanced Resource utilization:} As shown in Figure~\ref{fig:reource_utilization}, the DQN-Scheduler increases resource utilization for most PMs for both CPU and memory, whereas other algorithms consolidate tasks into fewer PMs. This indicates that the DQN-Scheduler achieves the best outcomes in maintaining resource utilization for most PMs within the proposed upper and lower thresholds. For example, the DQN-Scheduler's mean CPU utilization is 1.3043 cores and the standard deviation is 0.7499, indicating steady CPU utilization. Similarly, memory utilization has a mean value of 8.57927 GB with a standard deviation of 3.3464 (Table \ref{tab:algorithm_performance} and Figure \ref{fig:CPUandMemorySDandMean}). This emphasises that the DQN-Scheduler not only enhances resource utilization across all PMs but also boosts the overall performance of the cloud environment.

The small margin seen in the SD results in Table \ref{tab:algorithm_performance} establishes the DQN-Scheduler as the second-best performer across key metrics such as CPU utilization mean 1.3043 cores, memory utilization mean of 8.57927 GB, and latency mean of 4.8555. However, none of the current approaches consistently achieve top rankings across all metrics. Regarding resource utilization, GA has the lowest standard deviation in CPU utilization, just slightly better than the DQN-Scheduler by less than 0.01. GA achieved a mean CPU utilization of 1.299. However, GA has the highest standard deviation in memory utilization, with a value of 4.3192 and a mean of 9.073. This indicates that it performs worst in memory usage (Figure \ref{fig:CPUandMemorySDandMean}).

On the other hand, PSO has a lower standard deviation in memory utilization, with a value of 3.2842 and a mean of 8.60123. This indicates that PSO is consistent with memory utilization. However, PSO shows a standard deviation in CPU utilization of 1.00558 and a mean of 1.4432, indicating the worst for CPU usage.

This shows that the DQN-Scheduler efficiently balances resource utilization on each PM. In addition, its consistent performance across various resource metrics indicates that the DQN-Scheduler can effectively manage CPU and memory utilization. Overall, the DQN-Scheduler shows a robust performance across various resource metrics, suggesting its potential for enhancing the sustainability and stability of cloud systems.

\subsubsection{Quality of Service}
This section evaluates the performance of our proposed framework in terms of SLA violations, comparing it to benchmark algorithms as modelled in Section~\ref{system_model}.

The equation (\ref{EQ:9}) is proposed to calculate the SLA violation time per active host when hosts experience intensive CPU utilization. The equation considers that SLA violations occur when CPU utilization exceeds 100\%. However, since we limit utilization to an upper threshold, we consider SLA violations when CPU utilization exceeds our proposed upper limit of CPU.

\textbf{\underline{SLA violations:}} Over a month-long experiment, $T_{an}$ represents the total operational time of all microservices in the cluster. As a result, the GA algorithm had 0.000093 SLA violations during the experiment. The DQN-Scheduler did not violate SLA during the experiment.  This indicates the DQN-Scheduler utilizes the resources efficiently Table \ref{tab:algorithm_performance}.

\textbf{\underline{Reliability of the Scheduling Process:}}

Equations (\ref{EQ:9}) and (\ref{EQ:10}) employ the SD and mean of entropy to evaluate scheduler reliability. They reflect the variability and uniformity of microservice distribution across PMs. However, before diving into the results and analyzing them, the significance of the SD and the mean values are explained. There are four different combinations of the SD and the mean. Each combination has a specific interpretation. While entropy typically represents uncertainty and randomness in distribution \cite{amorocho1973entropy}, we use it to assess the variation of microservices in each PM.

\begin{figure}[h!]
    \centering
    \includegraphics[width=0.50\textwidth]{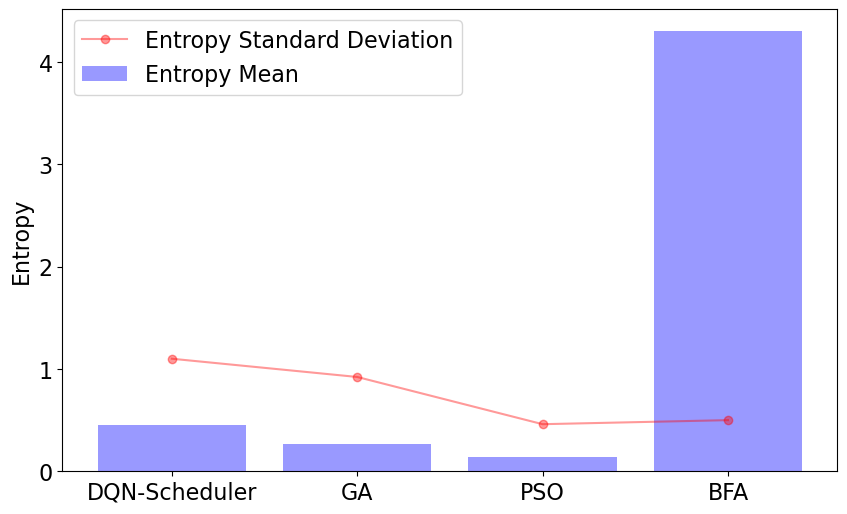}
    \caption{Microservices Variation and System Reliability Using Entropy: Standard Deviation and Mean}
    \label{fig:entropy}
\end{figure}

We expect that scheduling distributes microservices to different PMs to reduce the probability of a single failure of an application and enhance availability and reliability. The main goal is to increase the variability of microservices across PMs. The following explains the implications of the different combinations of the mean and the SD:
\begin{enumerate} \item 
\textbf{ Low Standard Deviation and High Mean Entropy:} It indicates that microservices are well-distributed among PMs with minimal clustering. 
\item \textbf{ High Standard Deviation and High Mean Entropy:} It indicates that microservices are generally distributed evenly across PMs, with notable variation in distribution.
    \item \textbf{Low Standard Deviation and Low Mean Entropy:} It indicates most microservices are clustered on a few PMs, showing consistent clustering.
    \item \textbf{High Standard Deviation and Low Mean Entropy :} It indicates a strong tendency for clustering, with significant variation among PMs. It also suggests a mixture of clustered and even distributed VMs. 

\end{enumerate}
The results in (Table \ref{tab:algorithm_performance} and Figure \ref{fig:entropy}) indicate that the DQN-Scheduler has the highest SD with a value of 1.1033 and a low mean with values of 0.4547. This suggests notable variability of microservices in each PM. This demonstrates that the DQN-Scheduler most effectively enhances the reliability of the cloud. GA results show the second-highest SD with a value of  0.9242 but a low mean entropy of 0.26059. The results indicate strong clustering, with some variation in microservices across PMs. The PSO and BFA have results for the SD, with values of 0.4615 and 0.4490, respectively. They have 0.1412 and  4.31443 values as the mean, respectively, suggesting less variability of microservices in PMs.Additionally, BFA has the highest mean entropy with a value of 4.3144 (Table \ref{tab:algorithm_performance} and Figure \ref{fig:entropy}). The results indicate minimal clustering of microservices. 

The DQN-Scheduler has a relatively low mean entropy and the highest SD, indicating a strong tendency for clustering with notable variation in microservices. GA has low mean entropy and the second-highest SD, suggesting some clustering among microservices distributed to PMs. The results indicate that GA follows our framework as the second-best reliable algorithm. BFA comes in third with the highest mean entropy but a lower SD, indicating minimal clustering and less variation. The PSO results indicate microservices are clustered on a few PMs with the least variation, implying a lower level of reliability.

In summary, the DQN-Scheduler has the best combination of mean and SD, demonstrating the most effective scheduling approach to enhance reliability.

\subsubsection{Latency}

In this section, we discuss the results of latency caused by communication between distributed microservices. Improving latency can be achieved by making microservices independent or by scheduling dependent microservices close to each other. However, while scheduling dependent microservices close to each other can reduce latency, it must be done cautiously to avoid causing system reliability. For example, clustering many replicates of a microservice in the same PM increases the risk of a single point of failure, thereby reducing reliability.
\begin{figure}[h!]
    \centering
    \includegraphics[width=0.485\textwidth]{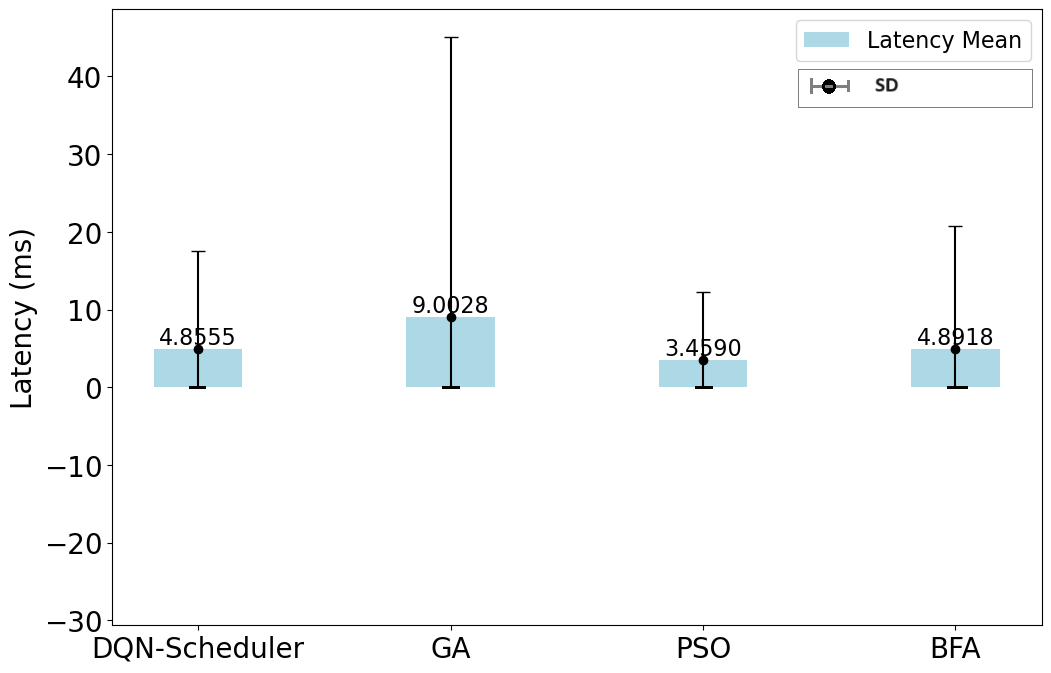}
    \caption{Latencies Caused by Approaches: Standard Deviation and Mean}
    \label{fig:latency}
\end{figure}

The results shown in Table \ref{tab:algorithm_performance} regarding latency demonstrate that the PSO algorithm has the lowest latency, with an SD of 8.8421 and a mean latency of 3.459 ms, compared to other approaches. These results suggest that most microservices have latency around this mean value. However, PSO clusters microservices in fewer PMs, reducing reliability. Therefore, it is expected that PSO has the lowest latency but also lower reliability (Table \ref{tab:algorithm_performance} and Figure \ref{fig:latency}).

On the other hand, the DQN-Scheduler exhibits the second-lowest latency, with an SD of 12.7412 and a mean latency of 4.8555 ms. This indicates a balance between reliability and latency. BFA ranks third, with an SD of 15.8433 and a mean latency of 4.8918 ms. The GA produces the highest latency, with an SD of 30.0443 and a mean of 9.0028 ms (Table \ref{tab:algorithm_performance} and Figure \ref{fig:latency}).

Overall, The DQN-Scheduler proved that it balances resource utilization, reliability, availability, and latency better compared to other approaches. It is the only scheduler that manages trade-offs between conflicting objectives and finds the best solutions. 

\subsection{Comparison with Current Studies}
\begin{table*}[h!]
\centering
\footnotesize
\begin{tabular}{|l|c|c|c|c|c|c|c|c|c|c|c|c|}
\hline
\textbf{Study} & \rotatebox{90}{\textbf{Response Time}} & \rotatebox{90}{\textbf{Load Balancing}} & \rotatebox{90}{\textbf{Scalability}}  & \rotatebox{90}{\textbf{Energy Consumption}}&\rotatebox{90}{\textbf{Availability}}  & \rotatebox{90}{\textbf{Resource Utilization}} & \rotatebox{90}{\textbf{Request Success Rate}} & \rotatebox{90}{\textbf{Latency}} & \rotatebox{90}{\textbf{Makespan}} &\rotatebox{90}{\textbf{QoS Satisfaction}}&\rotatebox{90}{\textbf{Real-time Scenario}}  \\
 
\hline
RSDQL\cite{lv2022microservice} & $\checkmark$ & &  & & && & &&&   \\
\hline
\cite{9723469} & $\checkmark$ & &  & $\checkmark$ & &$\checkmark$ & & &&&   \\
\hline
Noah \cite{10098822} & $\checkmark$ & &  & && $\checkmark$ & $\checkmark$ && &&   \\
\hline
GRLD\cite{10162207} & $\checkmark$ & &  & & && && &$\checkmark$&   \\
\hline
DQTS\cite{tong2020scheduling} & & $\checkmark$ & & && & && $\checkmark$ & &  \\
\hline
QEEC\cite{ding2020q} & $\checkmark$ &  & & $\checkmark$ && & && &&   \\
\hline
QL-HEFT\cite{tong2020ql} & $\checkmark$ & &  & && & && &&   \\
\hline
BCRN\cite{asghari2024bi} & && &  & &$\checkmark$ & & &$\checkmark$ & &   \\
\hline
DDQ-EES\cite{zhang2018double} & & & & $\checkmark$ &&& && &&   \\
\hline
DQN Framework \cite{peng2020multi} & &  & & $\checkmark$ & & & &&$\checkmark$  &&  \\
\hline
E-AEO-AOA\cite{yeganeh2023novel} & &  & & $\checkmark$ && & &&$\checkmark$  & &   \\
\hline
WDDQN-RL\cite{li2022weighted} & & & & & &$\checkmark$& & &$\checkmark$ && \\
\hline
\textbf{DQN-Scheduler} & &$\checkmark$ &$\checkmark$ & &$\checkmark$& $\checkmark$& & $\checkmark$& && $\checkmark$\\
\hline
\end{tabular}
\caption{Comparison of Various Studies on Different Metrics}
\label{tab:Comparison of Various Studies on Different Metrics}
\end{table*}
This section provides insight into the comparison between existing studies and our proposed framework. We found that current studies are divided into three categories based on the number of objectives: single-objective, bi-objective, and multi-objective (see Table \ref{tab:Comparison of Various Studies on Different Metrics}).
Single-objective studies, such as RSDQL \cite{lv2022microservice}, QL-HEFT \cite{tong2020ql}, and DDQ-EES \cite{zhang2018double}, do not aim to enhance overall cloud performance. For example, while RSDQL and QL-HEFT may effectively reduce response time, they intensively utilize resources and increase energy consumption. Similarly, DDQ-EES focuses solely on reducing energy consumption, ignoring QoS parameters such as makespan, reliability, latency, and load balancing. Thus, approaches that consider only one objective may achieve high results in a particular aspect without considering other factors that enhance overall performance. Focusing on one aspect of cloud scheduling can negatively affect other aspects.

The second type of study considers two objectives. Studies such as GRLD \cite{10162207}, DQTS \cite{tong2020scheduling}, QEEC \cite{ding2020q}, BCRN \cite{asghari2024bi}, the DQN Framework \cite{peng2020multi}, and E-AEO-AOA \cite{yeganeh2023novel} propose to address two non-conflicting objectives Table (\ref{tab:Comparison of Various Studies on Different Metrics}). These studies aim to enhance makespan along with load balancing, resource utilization, or energy consumption. However, they do not consider the cost of utilizing the network, latency, and reliability, which directly affect QoS and performance.
In all bi-objective approach models, the workload needed to be scheduled is considered the latency and network costs are not considered. This is due to:
\begin{enumerate}
\item During the experiment, the workload and cloud environment configurations were small-scale. Resulting in making latency and network traffic costs are negligible (e.g., DQTS and BCRN).
\item The assumption that workloads are scheduled at the same time on the same nodes (e.g., QEEC and BCRN).

\end{enumerate}

E-AEO-AOA \cite{yeganeh2023novel} aims to balance the load and enhance energy efficiency by offloading dependent tasks to fog computing. However, it ignores the cost of utilizing resources (e.g., CPU, memory, and network) and latency, which directly affect scheduling performance.

The third type of study, considering more than two objectives, comprises about 15\% of current research. However, these studies also have limitations in providing online and real-time scheduling. For example, \cite{9723469} trains their model using an offline approach that gathers data from past events, limiting the training process to certain environmental conditions. Similarly, Noah \cite{10098822} uses a synthetic dataset for training, which may handle unseen events in the Alibaba data center but challenges optimal decisions for other data centers.

To summarize, the multi-objective approaches currently suffer from:
\begin{enumerate}
    \item Generalizing the decision-making process for scheduling workloads.
\item Complex models that do not consider conflicting objectives.
  \item Ignoring real-time scheduling scenarios.
  \item Utilizing small-scale datasets for validating the proposed algorithm, lacking the reality of reflecting on model performance.
\end{enumerate}

As a result, our proposed framework enhances the scheduling process for microservices and outperforms existing studies. Our framework considers resource utilization, online scheduling, availability, reliability, latency, and real-world scheduling scenarios to enhance the overall performance of cloud computing.

\section{Conclusion and Open Question}  

Our framework aims to use an agent as part of other components to enhance the multi-objective scheduling process. The agent contributes to scheduling microservices efficiently. We tackled four conflicting objectives: resource utilization management, load balancing, reliability, and latency. Optimizing these objectives is extremely challenging; therefore, the agent was rewarded with utilizing the Pareto Front approach. The DQN-Scheduler has components to enhance the online scheduling process, such as resource management, SLA monitoring, and cluster management. The DQN-Scheduler utilizes an agent's action selection mechanism, utilizing Bayesian inference. We employ the mechanism to ensure the selection of the best actions in a real-time manner. 

The DQN-Scheduler enhances resource utilization by spreading microservices in groups across the cluster PMs. In addition, it balances the load of each PM's resources (CPU and memory). This results in improved resource utilization and enhances the long-term performance of the cloud platform. Moreover, the DQN-Scheduler improves the QoS of the cloud by complying with scheduling constraints. The trained agent and its awareness of the resource utilization threshold benefit our scheduling framework by managing the resource utilization of PMs. There was no incident where The DQN-Scheduler violated the SLA by utilizing CPUs extensively. Furthermore, the DQN-Scheduler has clustered microservices in each PM, with a notable variation of microservices within each cluster, leading to the highest level of reliability.

Finally, The DQN-Scheduler significantly reduces latency by clustering dependent microservices into the minimum number of PMs. Our DQN-Scheduler carefully balances reducing latency and enhancing the reliability and availability of microservices.

In summary, our proposed framework outperformed baseline algorithms in multiple metrics, demonstrating its ability to balance conflicting objectives, including resource utilization, load balancing, reliability, and latency. Future work could focus on optimizing microservice scaling while considering the increased complexity of the system.

% \ifCLASSOPTIONcaptionsoff
%   \newpage
% \fi
\bibliographystyle{IEEEtran}
\bibliography{references}
\end{document}